\documentclass[aip,apm,citeautoscript,reprint]{revtex4-2}

\usepackage{graphicx}
\usepackage{dcolumn}
\usepackage{bm}

\usepackage[utf8]{inputenc}
\usepackage[T1]{fontenc}
\usepackage{xcolor}
\usepackage{mathptmx}
\usepackage{amsmath}
\usepackage{etoolbox}

\makeatletter
\def\@email#1#2{%
 \endgroup
 \patchcmd{\titleblock@produce}
  {\frontmatter@RRAPformat}
  {\frontmatter@RRAPformat{\produce@RRAP{*#1\href{mailto:#2}{#2}}}\frontmatter@RRAPformat}
  {}{}
}%
\makeatother
\begin{document}

\preprint{AIP/123-QED}

\title[Quantum materials for energy-efficient neuromorphic computing]{Quantum materials for energy-efficient neuromorphic computing: Opportunities and challenges}

\author{Axel Hoffmann}
\affiliation{Materials Research Laboratory and Department of Materials Science and Engineering, University of Illinois Urbana-Champaign, Urbana, Illinois 61801, USA}

\author{Shriram Ramanathan}
\affiliation{School of Materials Engineering, Purdue University, West Lafayette, Indiana 47907, USA}

\author{Julie Grollier}
\affiliation{Unit\'e Mixte de Physique CNRS/Thales, Universit\'e Paris-Saclay, 91767 Palaiseau, France}

\author{Andrew D. Kent}
\affiliation{Center for Quantum Phenomena, Department of Physics, New York University, New York 10003, USA}

\author{Marcelo Rozenberg}
\affiliation{Universit\'e Paris-Saclay, CNRS Laboratoire de Physique des Solides, Orsay 91405, France}

\author{Ivan~K.~Schuller}
\affiliation{Department of Physics, University of California--San Diego, La Jolla, California 92093, USA}
\affiliation{Center for Advanced Nanoscience, University of California--San Diego, La Jolla, California 92093, USA}

\author{Oleg Shpyrko}
\affiliation{Department of Physics, University of California--San Diego, La Jolla, California 92093, USA}

\author{Robert Dynes}
\affiliation{Department of Physics, University of California--San Diego, La Jolla, California 92093, USA}

\author{Yeshaiahu Fainman}
\affiliation{Department of Electrical and Computer Engineering, University of California--San Diego, La Jolla, California 92093, USA}

\author{Alex Frano}
\affiliation{Department of Physics, University of California--San Diego, La Jolla, California 92093, USA}

\author{Eric~E.~Fullerton}
\affiliation{Center for Memory and Recording Research, University of California--San Diego, La Jolla, California 92093, USA}
\affiliation{Department of Electrical and Computer Engineering, University of California--San Diego, La Jolla, California 92093, USA}

\author{Giulia~Galli}
\affiliation{Pritzker School of Molecular Engineering and Department of Chemistry, The University of Chicago, Chicago Illinois 60637, USA}
\affiliation{Materials Science Division, Argonne National Laboratory, Lemont, Illinois 60439, USA}

\author{Vitaliy Lomakin}
\affiliation{Center for Memory and Recording Research, University of California--San Diego, La Jolla, California 92093, USA}
\affiliation{Department of Electrical and Computer Engineering, University of California--San Diego, La Jolla, California 92093, USA}

\author{Shyue Ping Ong}
\affiliation{Department of NanoEngineering, University of California--San Diego, La Jolla, California 92093, USA}

\author{Amanda K. Petford-Long}
\affiliation{Materials Science Division, Argonne National Laboratory, Lemont, Illinois 60439, USA}
\affiliation{Department of Materials Science and Engineering, Northwestern University, Evanston, Illinois 60208, USA}

\author{Jonathan A. Schuller}
\affiliation{Department of Electrical and Computer Engineering, University of California Santa Barbara, Santa Barbara, California 93106, USA}

\author{Mark D. Stiles}
\affiliation{Physical Measurement Laboratory, National Institute of Standards and Technology, Gaithersburg, Maryland 20899-6202, USA}

\author{Yayoi Takamura}
\affiliation{Department of Materials Science and Engineering, University of California, Davis, Davis, California 95616, USA}

\author{Yimei Zhu}
\affiliation{Department of Consdensed Matter Physics and Materials Science, Brookhaven National Laboratory, Upton, New York 11973, USA}



\date{\today}

\begin{abstract}
Neuromorphic computing approaches become increasingly important as we address future needs for efficiently processing massive amounts of data. The unique attributes of quantum materials can help address these needs by enabling new energy-efficient device concepts that implement neuromorphic ideas at the hardware level.  In particular, strong correlations give rise to highly non-linear responses, such as conductive phase transitions that can be harnessed for short and long-term plasticity.  Similarly, magnetization dynamics are strongly non-linear and can be utilized for data classification.  This paper discusses select examples of these approaches, and provides a perspective for the current opportunities and challenges for assembling quantum-material-based devices for neuromorphic functionalities into larger emergent complex network systems.
\end{abstract}

\maketitle

\section{\label{Sec:Introduction}Introduction}


Brain-inspired computing is promising for the development of highly efficient complex computation with minimal energy consumption.  Today, software implementations of neural networks have become ubiquitous, including facial recognition software, language translation, and search engines. Unfortunately the energy being used on these tasks is growing unsustainably. Fortunately, the brain also provides inspiration for approaches to minimize this energy. Brain-inspired computing can not only address the efficiency of computing in data centers, it can also address the need to provide efficient computation solutions locally at potential sensors and close to the end-user in order to avoid the inefficiencies of communicating data over long-distances. 

Most research on neural networks and on neuromorphic computing uses CMOS technology.\cite{furber2016large} However, as an increasing number of applications emerge, more complicated networks require more resources. One approach to limit the energy used is to use specialized hardware like graphical processing units (GPUs) or tensor processing units (TPUs) that are optimized for carrying out the many vector-matrix multiplications that are used in neural networks. So far, this is the most commercially viable approach. 

In a more forward-looking approach to capturing some of the brain's efficiency for cognitive computing, many researchers are trying to more closely mimic the brain by building spiking networks. Large corporations\cite{davies2018loihi,furber2012overview,merolla2014million} 
have developed specialized chips that are in some sense, still conventional digital computers, since they are implemented with standard CMOS technology but they are optimized for running artificial neural networks algorithms. These chips are designed around digital CMOS implementations of synapses and neurons, where a large number of relatively small computing units are in close physical proximity to the working memory that is distributed throughout the system. These remarkable chips implement millions of neurons and hundreds of millions of synapses.

Other approaches\cite{schemmel2012live} 
are hybrid in the sense that the neurons are physically implemented by analog electronics.\cite{thakur2018large} For instance, the membrane potential of each neuron is implemented by the charge stored in a capacitor. The neurons emit a spike when the voltage of the capacitor reaches a preset threshold potential. However, the system is hybrid because the spike is not represented by the emission of an analog action potential, but by a spike event, which is communicated to a downstream neurons through a network that digitally implements the synaptic connections.
The power consumption of spiking systems might be further optimized by using transistors working in the subthreshold regime.\cite{benjamin2014neurogrid} 
However, operating in this regime leads to more variability in the electronic response. 

Miniaturization of CMOS electronics is feasible down to the nanometer scale. However, all neuromorphic computing systems that represent a neuron in a compact manner, require a capacitor, whose miniaturization remains a serious challenge. 
The miniaturization of the capacitors is crucially limited by the dielectric constant of CMOS compatible materials.  It may be possible to overcome this problem with a radically different approach that rests on different physical principles for the implementation of artificial spiking neurons.

One way to reduce energy consumption in artificial neurons and other approaches to brain-inspired computing is to exploit the pronounced non-linear properties of quantum materials. The principal idea behind this energy efficiency is that in quantum materials a small electrical stimulus may produce a large response that can be electrical, mechanical, optical, magnetic, etc. through a material change of state. For example, several oxides exhibit a metal-insulator transition in which a small voltage or current can produce a large (several orders of magnitude) change in resistivity.\cite{imada1998metal} Furthermore, during the resistive switching, Joule heating provides an intrinsic physical mechanism for emulating short and long-term plasticity as well as spike-timing-dependent plasticity,\cite{seo2011analog} key mechanisms for learning, without external electric pulse control.  In parallel, recent work\cite{grollier2016spintronic,grollier2020neuromorphic} shows that magnetization dynamics provide a rich variety of nonlinear behavior that can be harnessed for classification.  New quantum materials offer novel pathways for manipulating such magnetization dynamics and giving rise to new functionalities important for neuromorphic computing such as analog memory. These properties then provide the material basis for emulating neurons, synapses, axons, and dendrites.


In a human brain, learning and memory are governed by synapses. One way to emulate synaptic behavior, is to base the artificial synapses on modulation of electrical resistivity by an electric stimulus (current or voltage). The nonvolatile synaptic “weights”, ({\em i.e.}, resistivity) must be set by an energy efficient method and they should vary continuously.  One traditional implementation of artificial synapses is based on resistive switching phenomena and uses phase change materials or transition metal oxides.\cite{zidan2018future,ielmini2021brain,wang2020resistive,lee2020nanoscale,xi2020memory} The continuous resistivity changes in phase change materials are created by intermediate phases. Density functional theory (DFT) calculations have shown that structural distortions result in a continuous change from crystallinity to amorphicity. Transition metal oxides are also used to emulate neurons (“neuristors”).\cite{choi2020emerging,yang2019memristive,yang2022nonlinearity} These are spiking devices, which exhibit leaky, integrate and fire behavior. The change in the volatile resistance from a high resistance state to low resistance state is caused by the application of a voltage higher than a threshold value, and from a low resistance to a high resistance state on removal of the applied voltage. The properties of these materials depend strongly on defects which introduce states in the energy gap changing the switching mechanism from thermal to electronic.


This paper focuses on identifying materials that can be used to engineer devices that perform as artifical neurons and synapses in a more compact manner than their CMOS equivalents. CMOS electronics has been designed for high precision numerical computing but have not been optimized for low precision categorical information like image processing or other targets of artificial intelligence. There are aspects of neural function in the brain, spike trains, oscillatory behavior in collections of neurons, that require significant chip area and/or energy when implemented in CMOS electronics.\cite{indiveri2011neuromorphic} The goal here is to facilitate such computation based on novel physics and devices that depend on material properties not found in doped silicon. Efficiency could be improved by using memristive crossbar arrays to enhance the efficiency of TPUs, creating more efficient spiking neurons or synapses, or by novel approaches that map to higher order neural activity.




One of the defining features of quantum materials is having complex Hamiltonians that yield unexpected and useful properties associated with various aspects of the collective electronic wave function – namely its lattice, spin, charge, and orbital character. This variety of phenomena offers some advantages over single-function semiconductors that are used most commonly in today’s devices. First, quantum materials can offer distinct functionalities built in the same material. For instance, a material can display resistive and magnetic properties that may be independently tunable with appropriate external perturbations. Second, the governing interactions, while microscopic in origin, thermodynamically couple across longer length scales in the material, yielding mesoscopic and macroscopic phases which are entirely emergent in nature. Thus, large networks based on quantum materials may be useful in neuromorphic computing because the emergent properties of the whole network (categorically distinct from the properties summing all individual devices) mimic the way the brain functions that are more complex than simply linearly combining neurons and synapses. In other words, the entire macroscopic state of the emergent neural network can retain information, learn, and potentially adapt to different stimuli. Central to these ideas is that quantum materials have properties governed by fundamental quantum mechanics at different length scales. 


Many quantum materials also harbor complex magnetic order or novel ways to connect charge transport to spin phenomena.  Towards the goal of new computational paradigms, magnetic spin torque oscillators can enable neuromorphic computing applications \cite{Torrejon2017neuromorphic}. In spin-torque oscillators, a spin-polarized current excites magnetization dynamics that can produce a high frequency electrical signal, typically in the range of 100~MHz to 50~GHz.  The resulting magnetization dynamics is nonlinear and tunable in phase, amplitude and frequency. Such oscillators are of great interest in neuromorphic computing where their response to external perturbations, phase locking and mutual synchronization can be exploited \cite{grollier2020neuromorphic}.  Coupled spin torque oscillators offer opportunities for complex neuromorphic functions.



Quantum materials enable new functionality in spin-torque oscillators by adding means of controlling and coupling such oscillators for neuromorphic computing. The new functionalities then can be controlled by the application of light, electric and magnetic fields.\cite{chakravarty_supervised_2019} Spin-torque oscillators can also be stochastic, ``jumping'' randomly between two or more discrete magnetic states.\cite{mizrahi2018neural} These characteristics can enable a compact and energy efficient source of random numbers,\cite{vodenicarevic2017low} including those needed to create stochastic binary networks for solving optimization problems.\cite{sutton2017intrinsic, borders2019integer}

The spin-torque oscillators studied to date are almost exclusively based on ferromagnetic metals. Examples include vortex oscillators or spin-Hall nano-oscillator arrays used to demonstrate neuromorphic computing.\cite{romera2018vowel,Zahedinejad2020,houshang2022phase} Vortex oscillators are composed of ferromagnetic metals that form a magnetic tunnel junction.\cite{dussaux2010large} A voltage bias on the junction causes a magnetic vortex in one of the electrodes of the junction to oscillate generating an oscillating electrical signal. Spin-Hall nanooscillators are formed from ferromagnetic nanoconstrictions in which a spin current excites spin waves in the nanoconstriction region that causes resistance oscillations.\cite{demidov2014nanoconstriction}

Quantum materials, such as transition metal oxides that exhibit phase transitions, can enable new oscillator functionalities. This is because the oscillator characteristics can change dramatically at phase transitions and, at a first order phase transition, can be hysteretic endowing the oscillator with memory, {\em i.e.}, its resonance frequency and output power can depend on its prior state \cite{zahedinejad2022memristive,Xu2021}. This can enable learning in a neuromorphic circuit, such as the adjustment of a synaptic weight. The exploration of these ideas has just begun.

Multiple oscillators can be coupled such that they phase lock, generating a larger output signal \cite{awad2017long}. Coupling can also be used as a means of propagating information in an array of oscillators. The coupling mechanisms can be electrical or magnetic. For example, if oscillators are connected in series, then the oscillating current generated by one oscillator will flow through the other oscillators providing a coupling mechanism.\cite{grollier2006synchronization} Oscillators have also been demonstrated to couple by spin waves when they “share” the same magnetic layer. Here spin waves emitted by one oscillator can be transmitted to another and have been shown to lead to frequency and phase locking.\cite{houshang2016spin} Quantum materials may offer additional coupling mechanisms through phase transitions that electrically couple oscillators.\cite{zahedinejad2022memristive,Xu2021} Tunable or even on/off coupling may be possible at magnetic phase transitions.

\begin{figure}[tb]
\centering
\includegraphics[width=3.2 in]{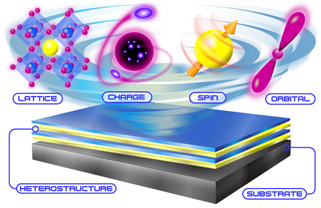}
\caption{\label{Fig:correlated}
Strongly correlated materials, specifically transition metal oxides, offer various degrees of freedom that can be tuned in heterostructures and can be useful for neuromorphic applications.
}
\end{figure}

\section{\label{Sec:NeuronSynapse}Neuron and Synapse Functionality}

\subsection{\label{Sec:MITCorrelated}Metal-insulator transitions in quantum materials}


Materials that exhibit resistive switching form the backbone of neuromorphic devices that circumvent traditional computing technologies based on CMOS logic and Von Neumann architectures.\cite{kuzum2013synaptic} Materials—such as VO$_2$—that exhibit weakly stimulated transitions between metallic and Mott insulating states represent the latest frontier in this research area. 
An important aspect that makes Mott insulators (and transition metal oxides in general) interesting materials for neuromorphic functionalities stems from the presence of valence band $d$-electrons. These electronic states exhibit characteristics between that of weakly localized $s$ and $p$ orbitals and highly localized, strongly magnetic, $f$-orbitals. Consequently, $d$\nobreakdash-electron states can undergo dramatic reorganizations due to changes in either the local ({\em e.g.}, carrier density, lattice structure, and interaction mechanisms) or global ({\em e.g.}, temperature, pressure and electric or magnetic fields) environment (see Fig.~\ref{Fig:correlated}). These reorganizations often significantly modify transport behavior, as exemplified by the several orders-of-magnitude change in resistivity seen across Mott metal-insulator transitions.\cite{mott1968metal}

\begin{figure}[tb]
\centering
\includegraphics[width=3.2 in]{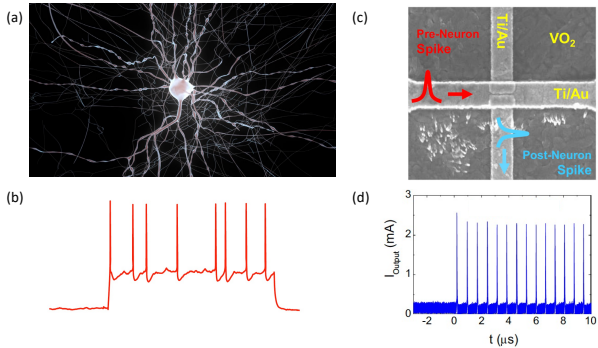}
\caption{\label{Fig:biological}
(a) An exemplary biological neuron that consists of dendrites, soma, and acon, with (b) a typical firing pattern triggered by neurotransmitters.  Adapted from Ref.~\onlinecite{markram2015reconstruction}.  (c) A quantum material neuristor composed by a VO$_2$ thin film and Au/Ti contacts.  (d)  Spiking dynamics triggered by Joule-heating across the metal-insulator transition in the neuristor.  Adapted from Ref.~\onlinecite{delvalle2020caloritronics}.
}
\end{figure}

One reason Mott insulators are of great interest in the context of artificial neuron devices is that recently, researchers showed that strong electric fields can induce electronic breakdown in these materials.\cite{kim2004mechanism} This resistive collapse can be induced by a succession of electric pulses analogous to the incoming electrical spikes that excite a brain’s neuron.\cite{delvalle2020caloritronics} In a Mott neuron device, electric excitations drive resistive collapse, leading to a sudden current surge that acts as an action potential spike. An example showing the similarities between biological neurons and artificial neuristors is shown in Fig.~\ref{Fig:biological}.  Importantly, this resistive switching is volatile:\cite{kim2004mechanism} the Mott material returns to a pristine insulating state after the applied voltage is terminated.  


While this behavior is experimentally well-established, the underlying mechanisms are still unclear. As established in the pioneering work of Ridley,\cite{ridley1963specific} the electric breakdown is related to negative differential resistance phenomena associated with the formation of conductive filaments. However, experimental evidence of filament formation during Mott insulator breakdown is mostly indirect and interpreted using numerical simulations. Understanding Mott electric breakdown is an important challenge for implementing practical and reliable artificial neurons. 
Since the Mott metal-insulator transition often coincides with a structural phase change, one should also explore local strain effects in filamentary structures, which may lead to slow relaxations. Interestingly, recent work by Del Valle {\em et al.}\ provides new experimental insight on the process of filamentary incubation \cite{delvalle2021spatiotemporal}.

Modifying the local defect concentration presents an appealing and effective new way of modifying the Mott metal-insulator transition. Examples include: controlling oxygen vacancies in vanadates,\cite{hu2014oxygen} doping rare-earth nickelates ({\em e.g.}, SmNiO$_3$) with light ions such as hydrogen,\cite{chen2019revealing} and controlling vacancies and doping in LaCoO$_3$.\cite{zhang2021predicting} Due to their high mobility, the distribution of light ions or ion vacancies is sensitive to short voltage pulses, enabling controllable tuning of neuromorphic device resistance. Memory nano-devices based on this effect are promising candidates for artificial neuromorphic synapses. 
Applying voltage pulses to hydrogen-doped perovskite ({\em e.g.}, $R$NiO$_3$, where $R$ is a rare earth cation) neuromorphic devices induces hydrogen dopant migration,\cite{chen2019revealing} enabling controllable tuning of electronic properties and new phase formation. Phase transformation has also been realized in La$_{0.7}$Sr$_{0.3}$CoO$_{3-\delta}$ —specifically, a series of topotactic transitions from the equilibrium ferromagnetic metallic perovskite structure ($\delta\approx 0$) to an antiferromagnetic  semiconducting brownmillerite structure ($\delta=0.5$) and further to a weakly ferromagnetic insulating Ruddlesden-Popper structure (La$_{1.4}$Sr$_{0.5}$Co$_{1+\nu}$O$_{4-\delta}$).\cite{chiu2021cation} A similar transition was also observed between the equilibrium brownmillerite SrCoO$_{2.5}$ phase to the metastable SrCoO$_3$ phase.\cite{lu2016voltage,jeen2013reversible,lu2017electric} Oxygen vacancy concentrations in the cobaltites have been varied in multiple ways, {\em e.g.}, by depositing oxygen-scavenging metals,\cite{gilbert2018ionic} annealing in reducing environments,\cite{jeen2013reversible} using electric fields,\cite{lu2017electric} and with epitaxial strain.

The existence of variable lifetime metastable states is another key aspect of quantum material phase transitions of great relevance to neuromorphic computing. In many cases the energy landscape features several accessible minima with different energies. Conversely, when an oxide is doped with charged ionic defects, the defect mobility allows for subtle changes in the material’s resistance. These changes, however, occur at slow timescales. This allows for metastable intermediate states that can be used to encode short term memory. The elasticity and plasticity of electrical resistance changes derived from tunable band-filling via mobile charged defects is a unique feature of strongly correlated oxides. This, in turn, enables the retention of information at various time scales, potentially mimicking a defining feature of the animal brain: memory storage and computational processes that occur at various time scales.

\paragraph*{Realizing synaptic functions in neuron systems.}

Quantum materials undergoing metal-insulator transitions, specifically VO$_2$, which has near room temperature metal-insulator transition,\cite{del2019subthreshold} have emerged among prominent candidates for neuristor platforms in bio-inspired devices. However, many fundamental questions related to the mechanisms of resistive switching behavior, stochastic electric response and the balance between volatile and non-volatile switching, remain an obstacle to the practical use of these devices in neuromorphic computing. 

\begin{figure}[t]
    \includegraphics[width=\columnwidth]{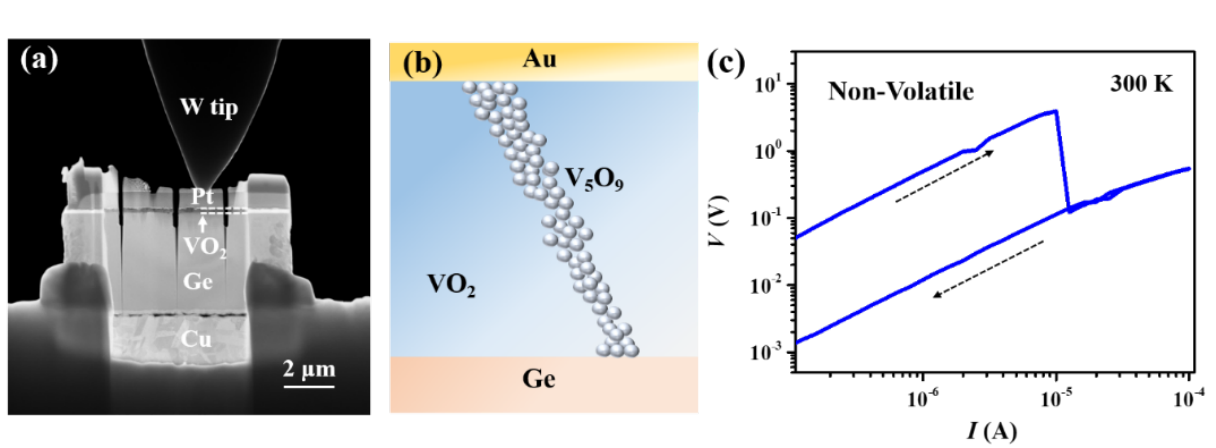}
    \caption{(a) A scanning transmission electron microscopy image showing the cross-section of a thin VO$_2$ device that can mimic the function of neuristor and synaptor for in-situ TEM experiments. (b) A schematic diagram showing the formation of V$_5$O$_9$ conductive filament in VO$_2$ matrix. (c) The non-volatile $I-V$ curve obtained at room temperature. From~\onlinecite{cheng2021operando}}
    \label{Fig:Microscopy}
\end{figure}

Realizing synaptic functions in neuron systems requires very-large-scale integration systems that contain electronic analog circuits to mimic neuro-biological architectures. The joints of these 3-dimensional cross-bar-shaped circuits are composed of numerous pre-neuristor lines, synaptor and post-neuristor lines. Naturally, different materials are are being considered to construct these operative lines in order to emulate the functions of neuristors and synaptors. As a result, the interfacial compatibility and strain effects of the materials will be a huge challenge for device fabrication. A single material that could realize both functions would be ideal. VO$_2$, because of its near room-temperature insulator-metal-transition and its subthreshold firing effects,\cite{del2019subthreshold} has been established as a good candidate for imitating the functions of neuristors.\cite{lin2016low} However, until now, no one has shown the use of VO$_2$ as synaptors. It is also possible to realize synaptic functions in VO$_2$, where the oxygen-vacancy containing V$_5$O$_9$ Magn{\'e}li phase acts as conductive filaments\cite{cheng2021operando} as demonstrated by combining in-situ transmission electron microscopy (TEM) and ex-situ transport measurements. Tuning the chemical composition, electric field, and working temperature gives a non-volatile switching process that can be removed by an annealing process, providing a “forget” function in synaptors, see Fig.~\ref{Fig:Microscopy}. 

These results illustrate that a single VO$_2$ material could provide the full function of a neuron cell.\cite{cheng2021operando} Recent studies have shown that it is possible to realize essential neuromorphic functions such as neurons, synapses and capacitors within a single nickelate device that can be reprogrammed on purpose by fast electric pulses.\cite{zhang2022reconfigurable}

\paragraph*{First-principles theoretical treatment of correlated electron systems.}

Recently, quantum-mechanical methods such as density functional theory (DFT) implemented in efficient codes 
have led to tremendous advances in predicting candidate materials for memory devices. Computational approaches can give insights into structural, electronic, and magnetic properties of broad classes of oxides and their associated memory technologies, including ferroelectric random-access-memory (FeRAM), phase-change memory, magnetic RAM (MRAM), and spin-transfer-torque MRAM (STT-MRAM). 
 
The structural and electronic transitions occurring in phase change materials can be impacted by defects.\cite{tran2021metal} Although difficult to unravel experimentally, the microscopic origin of these effects can be investigated computationally. For example, the lowering of vanadate metal-insulator transition temperatures due to point defects such as oxygen vacancies has been demonstrated computationally and also noted in experiments.\cite{tran2021metal} Rare earth nickelates, ReNiO$_3$, are other potential candidates for phase change materials. In these systems, Kotiuga {\em et al.}\ and {\em Cui et al.}\ respectively studied the effects of oxygen vacancies and Li doping in order to drive and understand changes in electronic structure that give rise to an improved memory device.\cite{kotiuga2019high,cui2021metal} A mechanistic understanding of the interplay between electronic structure changes and magnetic states was provided by computational investigations that could then be used to interpret experiments.  
 
Another set of phase change materials involving Sc-Sb-Te alloys were studied using first-principles simulations.\cite{wang2020time} The authors predict the structural evolution of these alloys upon optical excitation, providing a clear indication of the transition from a crystalline to an amorphous phase. Recently, perovskite oxides are gaining much interest as phase change materials for neuromorphic computing. The mechanism behind the metal insulator transition in perovskites has been addressed computationally for several systems, using first principles, DFT calculations. Zhang and Galli recently elucidated the impact of oxygen vacancies on the structural, electronic, and magnetic changes responsible for the metal-insulator transition in cobaltites.\cite{zhang2021predicting} Interestingly, they showed that cooperative structural distortions, rather than local bonding changes, are responsible for the gap closing. 
 
Zhang and Galli\cite{zhang2021predicting} also developed, and experimentally validated, a first-principles model that accurately predicts the electric bias required to drive the metal-insulator transition. Bennett et al.\cite{bennett2021origin} observed the effect of $n$-doping on the transition from FM metal to AFM insulatorin SrCoO$_3$. They showed that the metal-insulator transition is triggered by $n$-doping leading to a self-hole doped insulator, due to strong charge-lattice (or electron-phonon) coupling. As a result, controlling the hole-ligand ratio is a key factor in controlling the transition. Moreover, dopant metal ions and oxygen vacancies play a key role in non-volatile low power consuming memory devices; towards this end, clustering of oxygen vacancies around the Cu dopant in cubic $A$TiO$_3$ ($A$ = Ba, Be, and Mg) leads to the formation of conductive filaments that are responsible for electrical switching in such devices.\cite{resheed2021density} Similar mechanisms were also reported for $M$FeO$_3$ ($M$ = Gd, Nd) by substitutional doping of Al ions \cite{alsuwian2021first}.

Wang {\em et al.}\cite{wang2020time} studied the use of intermediate phases to realize synaptic behavior using DFT calculations; they showed that the structural distortions result in a continuous change from a crystalline to an amorphous state. Unlike phase-change-material-based synapses with continuously adjustable device conductances, phase-change-material-based neurons exhibit threshold firing. A change from high resistance state to low resistance state is only observed when the applied voltage is higher than a threshold value and returns to the high-resistive state when the applied voltage is removed. The effect of doping on the “firing’ of phase change material neurons has also been studied through first principles calculations.\cite{tran2021metal} For instance, introducing a defect instantly decreases the V$_2$O$_3$ band gap, switching the system from high resistance to low resistance.

\paragraph*{Challenges for first principles calculations.}

First principles calculations are useful tools to gain a microscopic understanding of transition metal oxides and predict the impact of doping, strain, or magnetic ordering on their electronic and atomic structure. These calculations are particularly useful at addressing these effects one at a time, which can be challenging to do experimentally, and at eventually understanding their complex interplay.  Although progress in DFT-based calculations has greatly accelerated our understanding of transition metal oxide properties, significant challenges remain. 

One challenge is to improve the accuracy of theoretical predictions of  strongly correlated materials with  highly localized $d$- or $f$-electrons. Standard semi-local DFT functionals are usually unable to treat such systems accurately. 
These problems may be mitigated to some extent by adding a semi-empirical Hubbard U parameter (DFT + U) to the Kohn-Sham Hamiltonians, using dynamical mean field theory, or using hybrid functionals. Semi-local functionals fail because of delocalization errors and self-interaction errors, which, in some cases, are partially fixed by the addition of a Hubbard U term to the Hamiltonian. The DFT+U method has yielded accurate results for the structural and electronic properties of several transition metal oxides.
However, appropriate U values are often determined empirically, and finding U values that reproduce properties across multiple phases of transition metal oxides can be challenging. 

Dynamical mean field theory, which has been applied both to realistic materials and to Hubbard models, has led to great progress in understanding correlated systems. Dynamical mean field theory provides an exact and accurate description of local intra-shell correlations. It was recently used\cite{singh2016selective} to understand correlation-driven charge order in the metallic VO$_2$ phase. 

Hybrid functionals, which incorporate a fraction of exact exchange, can sometimes yield more accurate structural and electronic properties of transition metal oxides than DFT+U descriptions, however they are much more demanding, from a computational standpoint. Wave function based methods, such as coupled-cluster, have been recently applied to transition metal oxides as well, showing improvement over DFT-based methods, although their use is still in its infancy. Finally we note that progress in understanding metal-insulator transition and other key properties of transition metal oxides at finite temperature is expected from coupling DFT-based or wavefunction based descriptions of the electronic structure with molecular dynamics.


\paragraph*{Enhancing device speeds via hidden phases.}

Current memristive devices typically operate with speeds in the 10~MHz to 350~MHz range. Achieving terahertz speeds would represent more than three orders of magnitude improvement over the fastest neuromorphic devices\cite{wang2020resistive} and lead to ultrafast processing of data and lower energy consumption. Hidden phases are material states that are typically not thermodynamically accessible, that is not reachable by bias field or temperature change, but can be activated by terahertz pulses.\cite{vaskivskyi2016fast,li2019terahertz} They emerge from out-of-equilibrium processes in a wide range of quantum materials, and could be key to building terahertz hardware synapses and neurons that require the emergence of multiple and controllable analog states. 

State-of-the-art large scale simulations\cite{prosandeev2021ultrafast} show that Pb(Mg$_{1/3}$Nb$_{2/3}$)O$_3$, the prototype of relaxor ferroelectrics, can emulate all the key neuronic and memristive synaptic features -- spiking, integration, and tunable multiple non-volatile states -- through terahertz pulse activation of hidden phases. The atomistic insight provided by these computations further reveals that the stabilization of multiple hidden phases occurs via the rearrangement, evolution and/or percolation of nanoscale regions in response to the terahertz electric field pulses. The different dielectric constants of the multiple hidden phases in relaxor ferroelectrics are also highly promising for creating ultrafast memcapacitor devices that are more energy-efficient than memristors, as no electric current runs through them during operation. As hidden phases occur in a broad class of quantum materials, including superconductors, colossal magnetoresistance manganites, charge-density wave materials, and incipient ferroelectrics, they open up a new avenue for creating quantum terahertz neuromorphic devices.

\paragraph*{Novel heterostructures for neuromorphic computing.}

Another unique aspect of strongly correlated oxides is the variety of macroscopic phases that they offer, sometimes within the same material or within two adjacently grown oxide materials (which has proven highly feasible using modern thin film deposition and heterostructuring tools). For instance, depending on the charge carrier concentration, a cuprate superconductor can also feature insulating antiferromagnetic properties.\cite{armitage2010progress} This phase diversity can be used to design arrays of Josephson junctions that exhibit neuromorphic behavior. Integrating a second material, such as a nickelate or vanadate, that displays neuromorphic behavior at different time scales or stimuli strengths would emulate the variety of scales seen in the behavior in the brain. Interfacing two materials can thus enable a combination of stimulus/response regimes. For instance, magnetic oscillator properties can be controlled via the tunable electrical resistance of an adjacent vanadate.\cite{Xu2021} Oxides may enable the combination of distinct charge, spin, or superconducting properties for future conversion of information modality from one form to another. 

\paragraph*{Engineering Mott materials for functionality.}

Although earliest implementations of Mott neuristors treated the underlying materials in a binary fashion,\cite{pickett2013scalable} either insulating or metallic, more recent studies have leveraged metastable and multi-state properties to achieve subthreshold firing.\cite{del2019subthreshold} The discovery of such unanticipated effects reflects one of the most interesting aspects of Mott neuromorphics: the complexity of metal-insulator transitions in Mott materials continues to yield surprising discoveries, portending future possibilities for device-level phenomena. For instance, the exquisite sensitivity of Mott insulators to small changes in processing parameter,\cite{zhang2017evolution} suggests powerful routes for engineering and controlling neuromorphic functionality through fine-tuning material composition and structure, applied strain, or light-induced phenomena.


\subsection{\label{Sec:MagnetizationDynamics}Magnetization Dynamics}

Magnetization states in magnetically ordered materials have a long history of being used for storing both digital and analog information.\cite{apalkov2016magnetoresistive,stamps20142014,dee2008magnetic} At the same time magnetic materials have several attributes that make them interesting for new neuromorphic computational schemes.\cite{locatelli2014spin}  The stability of magnetization states can be engineered with well defined variable relaxation times that can exceed years.  At the same time these magnetization states can still be manipulated with very low power, either through local magnetic fields or electrical means.\cite{wolf2003scanning,vzutic2004spintronics,Brataas2012,ohno2016spintronics}  This tunable non-volatility can be used for emulating synaptic weights. Conversely, it is possible to excite both stochastic and coherent magnetization dynamics.  In both cases the magnetization dynamics is determined by non-linear equations of motion, which can be used for thresholding behavior or other complex dynamic interactions that may resemble the functionality of natural neurons.  Research on both of these aspects has benefited significantly from the development of spintronics,\cite{wolf2003scanning,vzutic2004spintronics,Brataas2012,ohno2016spintronics} which is based on using the spin degree of electronic charge currents that allows to manipulate electronic charge currents through magnetization states and vice versa. In this section, we describe current-based means of manipulating magnetizations, the search for materials to optimize these effects, and some of the approaches to neuromorphic computing that can be done with them.\cite{grollier2016spintronic,sengupta2018neuromorphic,grollier2020neuromorphic}

\begin{figure}[tb]
\centering
\includegraphics[width=3.2 in]{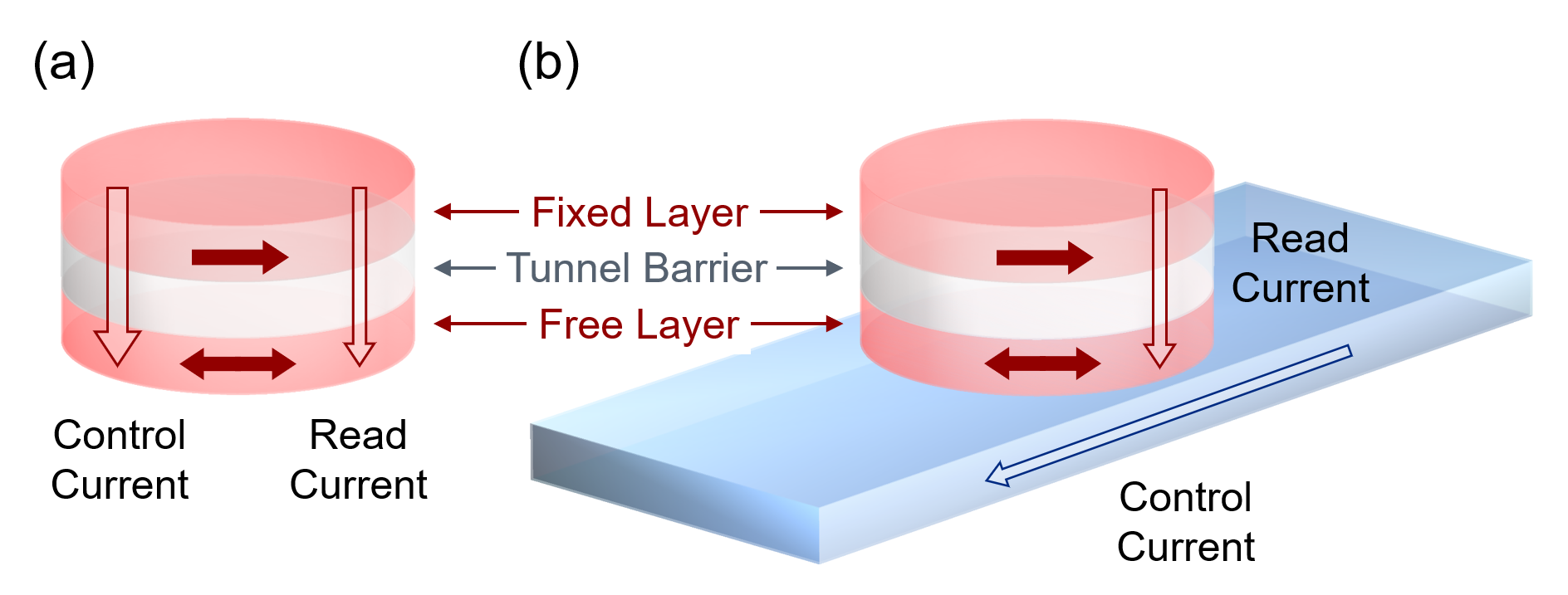}
\caption{\label{Fig:schematic}
Magnetic tunnel junctions (dark red arrows represent magnetization direction). (a) Standard magnetic tunnel junction with fixed and free layers and the control current following the same path as the read current. (b) Magnetic tunnel junction grown on the heavy metal layer with separate read and control current paths. From Ref.~\onlinecite{amin2020interfacial}.
}
\end{figure}

Two powerful ways to control magnetization dynamics are through spin-transfer torques\cite{stiles2006spin,ralph2008spin} and spin-orbit torques,\cite{manchon2019current} see Fig.~\ref{Fig:schematic}. Spin-transfer torques occur in multilayer devices with currents flowing perpendicular to the layers. The current passing through one magnetic layer becomes spin polarized and exerts a torque on the magnetization in a subsequent layer. This torque forms the basis for switching the memory states in magnetic random access memory (MRAM).\cite{Kent2015,apalkov2016magnetoresistive} It also leads to the dynamics in spin-torque nano-oscillators.\cite{silva2008developments,chen2016spin} Spin-transfer torques and spin-orbit torques are related, but in the latter case, the spin currents that create the torques arise from a current through an adjacent layer of a material with strong spin-orbit coupling. Each of these torques has its own disadvantages and advantages for controlling magnetization dynamics in spintronics-based neuromorphic applications.

One of the disadvantages of spin-orbit torques is the orientation of the torques. Both spin-transfer torques and spin-orbit torques are defined with respect to a spin direction. For spin-transfer torques, that direction is set by the direction of the magnetization in the polarizing layer and is relatively easy to control. For the dominant spin-orbit torques, the spin direction is set by geometry, being perpendicular to both the multilayer growth direction and the applied electric field. Such torques are very efficient at switching the magnetization in this same direction ({\em i.e.}, in the spin direction) but not so efficient for magnetizations that are in the growth direction, an orientation presently favored in MRAM applications.\cite{Kent2015,apalkov2016magnetoresistive} An active research direction is to efficiently produce torques with different spin directions. Doing so requires reducing the symmetry of the system. There are many ways to do so. One of the earliest is the use of spin-orbit layers with lower crystalline symmetry\cite{macneill2017control} but this approach has yet to produce torques strong enough to switch layers with perpendicular magnetizations. Alternatively, the magnetization in additional layers reduces the symmetry and can lead to perpendicular torques strong enough to reverse the magnetization.\cite{taniguchi2015spin,seung2018spin} This magnetization can be ferromagnetic\cite{taniguchi2015spin,seung2018spin} or antiferromagnetic.\cite{hu2021efficient} In both cases, it may be necessary to introduce a non-magnetic layer between the two magnetic layers to break the exchange coupling between them.

Interfaces\cite{hellman2017interface} play a crucial role in determining the strength of the spin-orbit torques in all possible mechanisms. In the  models in which the spin current is created in the interior of the generating layer,\cite{ando2008electric,liu2012current} the ferromagnetic exchange interaction at the interface converts the spin current into a torque. In another mechanism,\cite{manchon2008theory,miron2011perpendicular} the applied electric field combines with the spin-orbit coupling at the interface to create a spin accumulation at the interface that is misaligned from the magnetization and exerts a torque on it. Finally, in spin-orbit torque systems with a second magnetic layer, the interface of that layer can produce spin currents that lead to torques with useful directions through the processes of spin filtering and spin precession.\cite{amin2020interfacial}

\begin{figure}[tb]
\centering
\includegraphics[width=3.2 in]{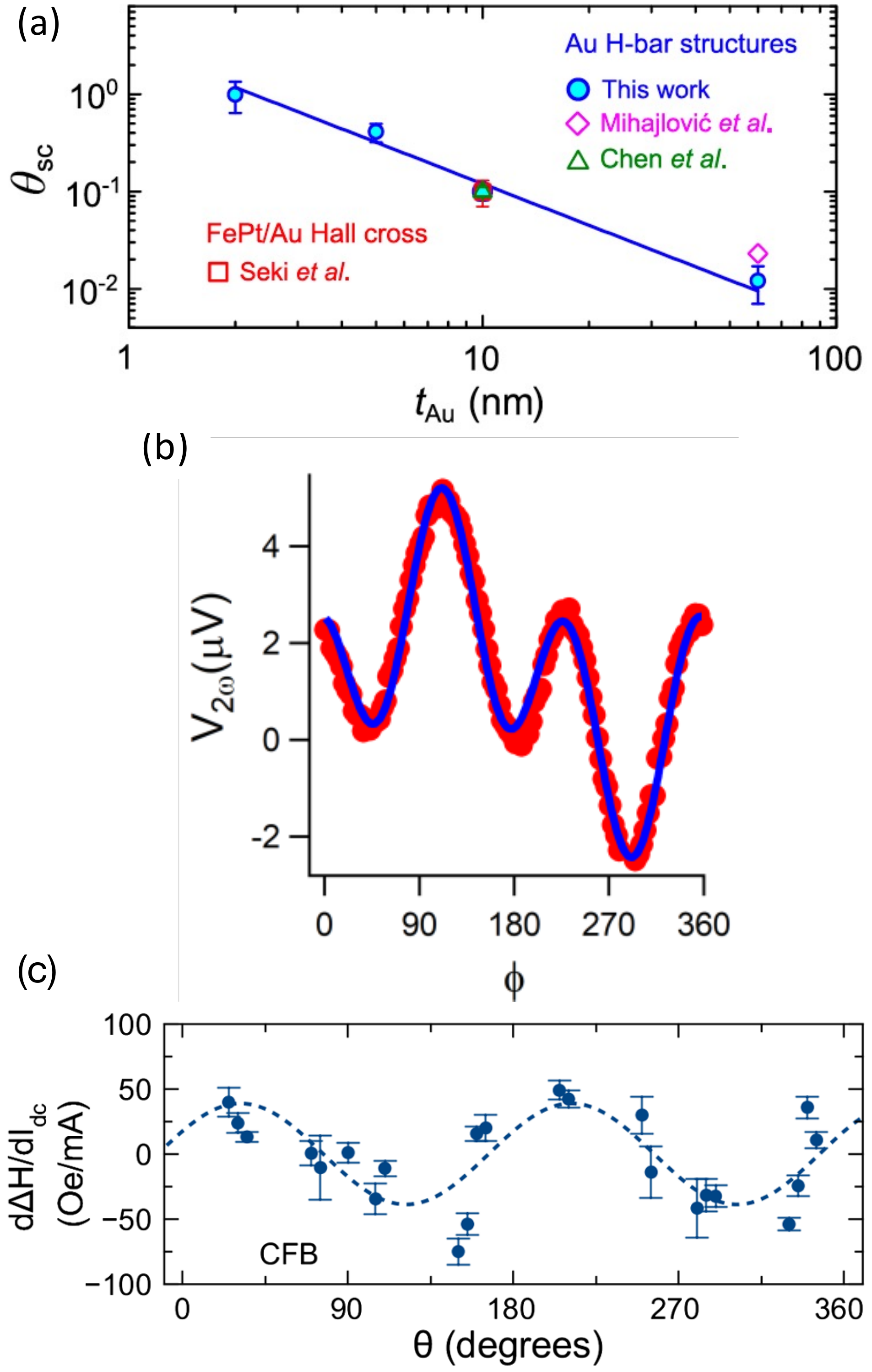}
\caption{\label{Fig:SOT}
New directions for spin-orbit torques.  (a) Enhanced spin-orbit torques from interfaces; spin to charge conversion measured in Si/Au multilayers through non-local transport as function of Au layer thickness compared to previous measurements of single layer Au films. The data for the single layers is from Ref.~\onlinecite{seki2008giant} for Seki {\em et al.}, Ref.~\onlinecite{chen2019generation} for Chen {\em et al.}, and Ref.~\onlinecite{mihaijlovic2009negative} for Mihajlovi\'{c} {\em et al.}  Adapted from Ref.~\onlinecite{elhadri2021large}.  (b) Unconventional torques due to magnetic symmetry breaking; second harmonic Hall measurements in FeRh/Ni$_{80}$Fe$_{20}$ bilayers indicate an unusual polarization direction of current induced spin accumulations in FeRh.  The polarization direction is about 45$^\circ$ with the current direction, which coincides with the equilibrium direction for antiferromagnetic spins in FeRh.  Adapted from Ref.~\onlinecite{gibbons2021large}. (c) Torques with the symmetry of the planar Hall effect associated with Co/Ni acting on a CoFeB layer. The variation of the spin-torque ferromagnetic resonance linewidth with dc current is plotted versus the magnetization angle $\theta$. The $\sin 2\theta$ variation is distinct from the symmetry of the spin-Hall effect. With a spin torque of this form the magnetization direction sets the spin-polarization direction {\em and} flow direction. Adapted from Ref.~\onlinecite{Safranski2020}.
}
\end{figure}

The quest for identifying new materials and novel heterostructures with efficient spin-orbit torques is still wide open.\cite{shao2021roadmap}  As mentioned above both interfaces and symmetry properties play an important role, which we would like to illustrate with three recent examples.  One of the early materials that were investigated for possible spin generation from charge currents was gold, but early measurements of the conversion efficiencies in gold thin films provided results that differed by at least one order of magnitude.\cite{seki2008giant,mihaijlovic2009negative}  A decade later an explanation for this possible discrepancy was provided by suggesting that the thickness of the gold films is very important and the charge to spin conversion can be significantly enhanced in thin films.\cite{chen2019generation}  However reducing the film thickness is limited by the requirement of having continuous films for transport.  It was shown that this limitation can be circumvented by using Si/Au multilayers,\cite{elhadri2021large} where the gold thickness can be reduced down to the nanometer-regime, which results in spin-charge conversion efficiencies or order unity, see Fig.~\ref{Fig:SOT}(a).  This is one example that shows that interfacial effects have the potential for significantly enhancing spin-orbit torques.

The second example for generating unconventional spin-orbit torques from low symmetries exploits that fact that the magnetic structure of antiferromagnetic materials can significantly reduce the symmetry that is given by the crystalline structure alone.  At the same time due to their vanishing net magnetization, their magnetic structure is typically robust against moderately strong magnetic fields, a fact that is advantageous for systems using spin torque oscillators that often require additional external magnetic fields. Exotic torques have been experimentally observed\cite{yang2019current,holanda2020magnetic,nan2020controlling} and theoretically predicted\cite{Mokrousov2013anisotropy,Freimuth2010anisotropic}.  A particular interesting case is provided by the antiferromagnet FeRh.  Here spin orbit torque efficiencies of up to 300~\% were observed at low temperatures, while angular dependent measurements also indicate that the polarization orientation of the electric current induced spins is determined by the direction of the spins in the antiferromagnet, see Fig.~\ref{Fig:SOT}(b). Thus antiferromagnets may provide a very efficient way to generate spin torques with geometries that are ordinarily unattainable.

The final example uses the result that ferromagnetic materials can also produce spin-orbit torques with novel symmetries.\cite{taniguchi2015spin} Spin-orbit coupling within ferromagnetic metals can be large and produce the well-known anomalous and planar Hall effects. The former, the anomalous Hall effect, can lead to spin currents polarized along the magnetization direction that flow perpendicular to the magnetization and electric field. Torques with the symmetry of the planar Hall effect are again polarized along the magnetization direction but flow parallel to the magnetization direction.\cite{taniguchi2015spin} Torques of this symmetry from a Co/Ni multilayer have been observed to act on a CoFeB free layer [Fig.~\ref{Fig:SOT}(c)]\cite{Safranski2020} and are the same order of magnitude as torques associated with the spin-Hall effect in Pt.\cite{Zhu2021} These torques may enable exciting spin waves and switching perpendicularly magnetized layers as well as permit new geometries for spin oscillators for neuromorphic computing.

Spintronic synapses harness the non-volatility of magnetization at the nanoscale to implement synaptic weights.\cite{GrollierIEEE2016} Several devices have been proposed with various degrees of biological resemblance. Spin-torque MRAM can store 32-bit floating point synaptic weights in ultra-low power chips.\cite{Ma-Endoh-2016} Magnetic tunnel junctions emulate binary weights by switching between two magnetization states.\cite{Greenberg-Toledo-Kvatinsky2021,jung2022crossbar,goodwill2021implementation} The intrinsic stochastic nature of magnetization switching in two-state magnetic tunnel junctions can also be leveraged for learning \cite{Vincent-Querlioz-2015}. Memristive behavior is obtained by modifying the magnetization texture to obtain gradual switching via spin-torque\cite{Lequeux-Grollier-2016,Sharad-Roy-2012} or spin-orbit torques.\cite{Kurenkov-Ohno-2019} Domain walls \cite{Zhang-Zhao-2021} or skyrmions \cite{Song-Woo-2020,Moutafis2020} imitate neurostransmitter nucleation and propagation in these structures. Optical excitations\cite{chakravarty_supervised_2019} and antiferromagnetic dynamics\cite{khymyn2018ultra} open the path to ultrafast, terahertz, spintronic synaptic devices.

Spintronic neurons leverage the non-linearity of magnetization dynamics. Reservoir computers based on large amplitude excitations of skyrmions,\cite{Pinna2020} domain walls,\cite{ababei_neuromorphic_2021} spin-waves,\cite{watt_implementing_2021} and vortices\cite{Torrejon2017neuromorphic,tsunegi_physical_2019} have been proposed. Related experiments typically rely on time-multiplexing the dynamics of a single magnetic dot or junction, and demonstrate state-of-the-art performance on tasks such as spoken digit recognition or memory capacity.\cite{Torrejon2017neuromorphic,watt_implementing_2021, tsunegi_evaluation_2018} Spintronic oscillators emulate the rhythmic features and synchronization behavior of biological neurons. Spin-torque and spin-orbit torque nano-oscillators solve classification tasks such as vowel recognition by synchronizing to external signals.\cite{romera2018vowel,Zahedinejad2020}. Large scale implementations of neural networks based on spintronic oscillators will require scaling down the oscillators, achieving mutual synchronization in large arrays, and controlling the synchronization process via dedicated synapses.\cite{prasad2021associative} 

\begin{figure}[t]
    \includegraphics[width=\columnwidth]{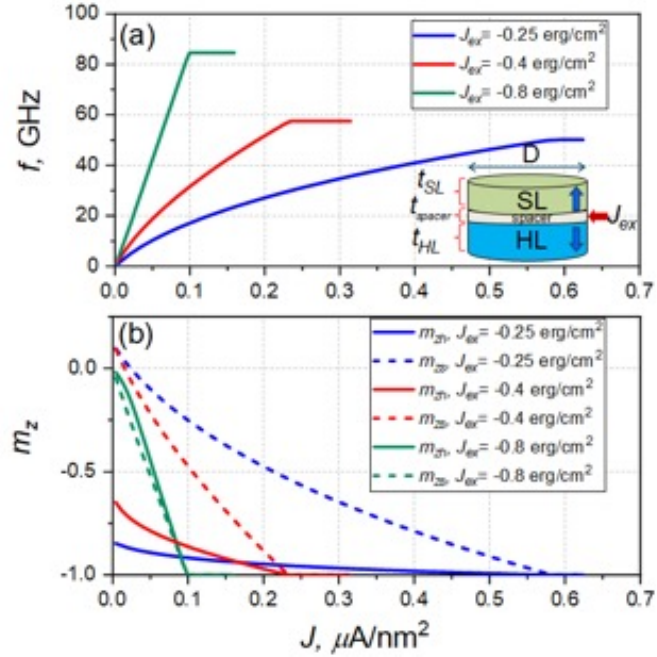}
    \caption{\textbf{Antiferromagnetic exchange coupled composite structure.} The inset shows the antiferromagnetically exchange coupled composite spin-torque oscillator structure, including a soft layer (SL), hard layer (HL), and spacer, {\em e.g.}, Ru layer for antiferromagnetic coupling. A perpendicular polarization layer is assumed to reside under the hard layer. The results obtained via the 2-macrospin model are shown for (a) precessional frequency   and (b) z-component of the magnetization in the soft layer and the hard layer as functions of the current density   for different surface exchange energy density coupling $J_\mathrm{ex}$ for $D=20$ nm, $t_h=t_s=0.8$ nm, $t_\mathrm{spacer}=0.3$ nm, $M_{s,s}=13.5\times 10^5$~A/m, $M_{s,h}=4.7\times 10^5$~A/m, $K_{u,h}=0.4$ MJ/m$^3$, $\alpha_h=\alpha_s=0.008$.}
    \label{Fig:VL1}
\end{figure}

Typical spin-torque-oscillator designs based on a single free layer require an external static magnetic field to operate but it is possible to eliminate the need for an applied field. One approach is based on an antiferromagnetically exchange coupled composite free layer.\cite{volvach2021micromagnetic} The mechanism of operation is based on the exchange field due to the antiferromagnetic coupling between the soft and hard sub-layers of the free layer. Such oscillators have large amplitude magnetization oscillations in the soft sub-layer that are tunable over a broad frequency range. They can generate an electric signal by placing a pinned in-plane layer with a magnetic tunnel junction or generate magnetic field outside the spin-torque oscillator. Figure~\ref{Fig:VL1} summarizes the operation of an antiferromagnetically exchange coupled composite spin-torque oscillator using a two-macrospin model. Micromagnetic modeling shows that for weaker currents, the two-macrospin model captures the behavior of the full model but for stronger currents the precession may become non-uniform depending on the oscillator size.

Electric field effects are particularly promising for controlling spintronic synapses as they are low power, can lead to non-volatile variations, and modify spin textures in multiple ways, {\em e.g.}, via interfacial effects in oxide/magnetic structures\cite{Xu2021,Fulara2020} or by locally changing the perpendicular anisotropy of the magnetic thin film. A future challenge is to go beyond the neighbor to neighbor coupling intrinsically delivered by exchange or dipolar coupling between the oscillator-neurons, and achieve controllable long range connections. It will also be interesting to design neurons with spiking behaviors as in biology, as well as increasing their speed above the gigahertz range. For this, antiferromagnets as well as synthetic antiferromagnets are promising due to their terahertz speed and spiking ability.\cite{Khymyn2017, markovic_easy-plane_2021} 

\begin{figure}[t]
    \includegraphics[width=\columnwidth]{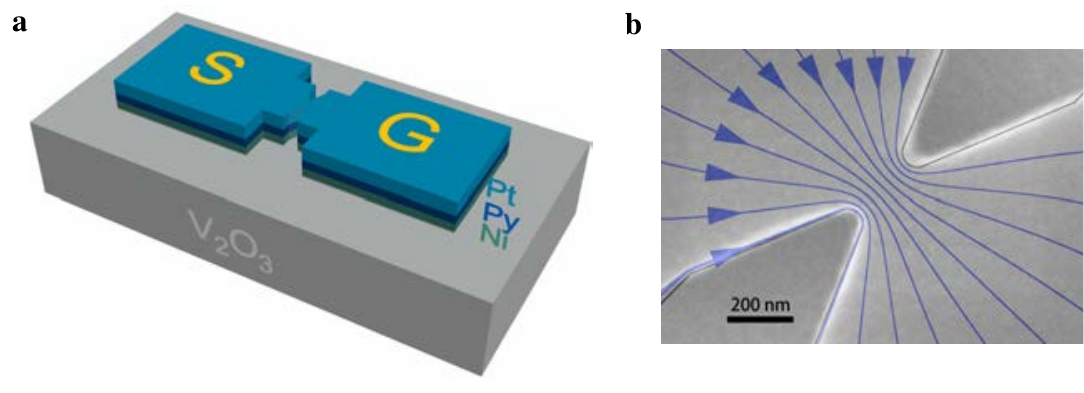}
    \caption{a) Schematic showing a hybrid metal‐insulator‐transition oxide/ ferromagnetic metal nanoconstriction. b) When the V$_2$O$_3$ is in an insulating state current is concentrated in the nanoconstriction region and excites spin waves that convert a rf current near the ferromagnet's resonance frequency to a dc voltage. The resonance frequency can be controlled with dc current through the constriction. From Ref.~\onlinecite{Xu2021}.}
    \label{Fig:QM_Osc}
\end{figure}

While new materials could provide efficient generation of magnetization dynamics as the basis for neuromorphic computing using spin oscillators \cite{Torrejon2017neuromorphic,romera2018vowel,Leroux2021radio,leroux2021hardware}, they do not address adjusting synaptic weights or enabling high connectivity between individual oscillators.  A neuromorphic circuit requires the output of a neuron to serve as the input to other neurons with learning associated with adjustment of the “synaptic” weights of the inputs to the neurons. High connectivity is also required, and an advantage of rf signal transmission between neurons.\cite{Leroux2021radio} This connectivity, however, requires transforming the rf output of a spin oscillator neuron back to a dc signal to serve as an input to other neurons. A spin resonator can provide this function. A rf input can excite spin precession that mixes the rf signal down to dc, known as the spin-diode effect\cite{tulapurkar2005spin} and spin-transfer ferromagnetic resonance.\cite{sankey2006spin-transfer} This behavior is common to spin resonators composed of ferromagnetic metals. However, being able to adjust the output characteristics is not. Incorporating a quantum material –– a metal‐insulator‐transition metal oxide –– can give spin resonators hysteresis and memory of their prior state,\cite{Xu2021} an important characteristic for oscillator-based synapses (Fig.~\ref{Fig:QM_Osc}). Specifically, it was shown that a dc current that heats the quantum material close to the metal insulator transition can be used to adjust the synaptic ``weight'' of the dc output signal.

This initial demonstration raises the question of what other materials may be well suited to integrate synaptic functionalities into spin oscillators.  Although much of the research on oxides over the last few decades has concentrated on the 3d transition elements (due to the smaller spatial extent of the wavefunction resulting in strong correlations), a few promising 5d oxides have recently emerged as useful tools in spin-based technologies because of the stronger spin orbit coupling of the 5$d$ electrons as compared to the 3$d$ electrons. various reports have shown that growing very thin layers of ruthenates,\cite{kan2016tuning,boschker2019ferromagnetism} iridates,\cite{matsuno2016engineering} or combinations of the two\cite{matsuno2016interface,ohuchi2018electric,hao2018giant} can result in new interfacial magnetic properties that can also be controlled with electric fields.\cite{ohuchi2018electric} The attractive aspect of these materials is that with the stronger spin orbit coupling, combined with the complex and tunable properties of oxides, a new platform for spin-based neuromorphic oscillators can be created which could feature intriguing magneto-transport properties in itinerant magnets with more energy favorable conditions to drive the oscillators.

The microwave properties of spintronic devices provide a unique perspective for supplying distributed power without hardwired connections. Neuromorphic systems reduce the energy consumption of neural networks by placing memory and computing devices as close to each other as possible and connecting them densely in a way that resembles the brain. Solving state of the art automatic classification tasks such as image recognition will require hundreds of millions of synaptic and neuronic devices. It is a challenge to bring the required power to each of these devices even if the individual energy consumption is low, due to the wiring complexity. Furthermore, the advantage of merging computing and memory may be lost if the inputs to be classified by the network need to be pre-processed to meet the requirements of a specific hardware. For example, the digitization of fast, analog signals and their transfer to a hardware neural network via a digital bus can consume several hundreds of watts. 

Spintronic devices are promising to solve these two issues because they can intrinsically sense analog signals and harvest the energy brought by these signals. The magnetization of nanoscale dots is highly sensitive to environmental electromagnetic fields that can be used as inputs to a neural network or to power the devices to which these dots are connected. Spin-diodes that convert radio-frequency signals to direct voltages are a great example of such process.\cite{tulapurkar2005spin} They can sense\cite{safin2020electrically} and harvest\cite{finocchio2021perspectives} microwave signals, and embody synaptic devices that perform the multiply-and-accumulate operations directly on input radio-frequency signals without digitization.\cite{leroux2021hardware} Interfaces with phase change materials control their synaptic weights directly.\cite{Xu2021}

In addition to microwave signals, spintronic devices are also sensitive to a wide range of signals such as optical beams and acoustic waves.\cite{hirohata2020review} A future direction of research for neuromorphic spintronics is thus to build ultra low-power neural networks that natively sense input signals and harvest the energy they need to operate.

\section{\label{Sec:Networks}Neural Networks}

\subsection{\label{Sec:ChargeCurrent}Charge Currents}


\paragraph*{Neurons and synapses, volatile and non-volatile resistive switching.}

As we have seen in previous sections, a great deal of progress has been achieved towards implementing artificial neurons and synapses that exploit the physical phenomena of resistive switching, also known as memristance. It is useful to recall that the artificial neuron and the artificial synapse rely on qualitatively different types of resistive switching, volatile and non-volatile, respectively. Biological neurons fire electric spikes but after those events they return spontaneously to a resting state or a quiescent state. Biological synapses, in contrast, modify their state, the synaptic coupling, during learning and then keep that state for a long time, even a lifetime as for significant memories. Thus, these two qualitatively different neural functions have a correlate in the volatile and non-volatile phenomena observed in resistive switching as mentioned above. We know from the architecture of the brain that its cognitive functions emerge from the interaction of a massive number of spiking neurons whose interconnections are modulated by synapses. Thus, the next stage in the development of neuromorphic hardware is to build networks of artificial neurons and synapses, which are subject to and evolve according to those two types of physical phenomena.

\paragraph*{An ideal neuromorphic quantum material.}

It is desirable to find a quantum material that can embody both types of resistive switching. A first step in that direction was recently demonstrated by Cheng {\em et al.}\cite{cheng2021operando} Those authors reported the non-volatile resistive switching of VO$_2$, which is a compound that exhibits volatile resistive switching near room temperature  associated with its metal-insulator transition.\cite{yi2018biological} The non-volatile behavior of VO$_2$ that was observed resulted from a chemical phase change (topotactic) transition into conductive V$_3$O$_5$ induced by a strong electric field, which caused the loss of oxygen ions. The transition could be reversed by externally heating the sample, making it interesting for applications, to achieve the recovery of VO$_2$ solely by electric means, such as self-heating. Other materials, which are worth investigating include the nickelates, such as SmNiO$_3$, which also display a metal-insulator transition driven by temperature,\cite{zhang2020organismic} making it potentially useful for implementing neurons via volatile switching. 

\paragraph*{Implementing Networks.}

\begin{figure}[t]
    \includegraphics[width=\columnwidth]{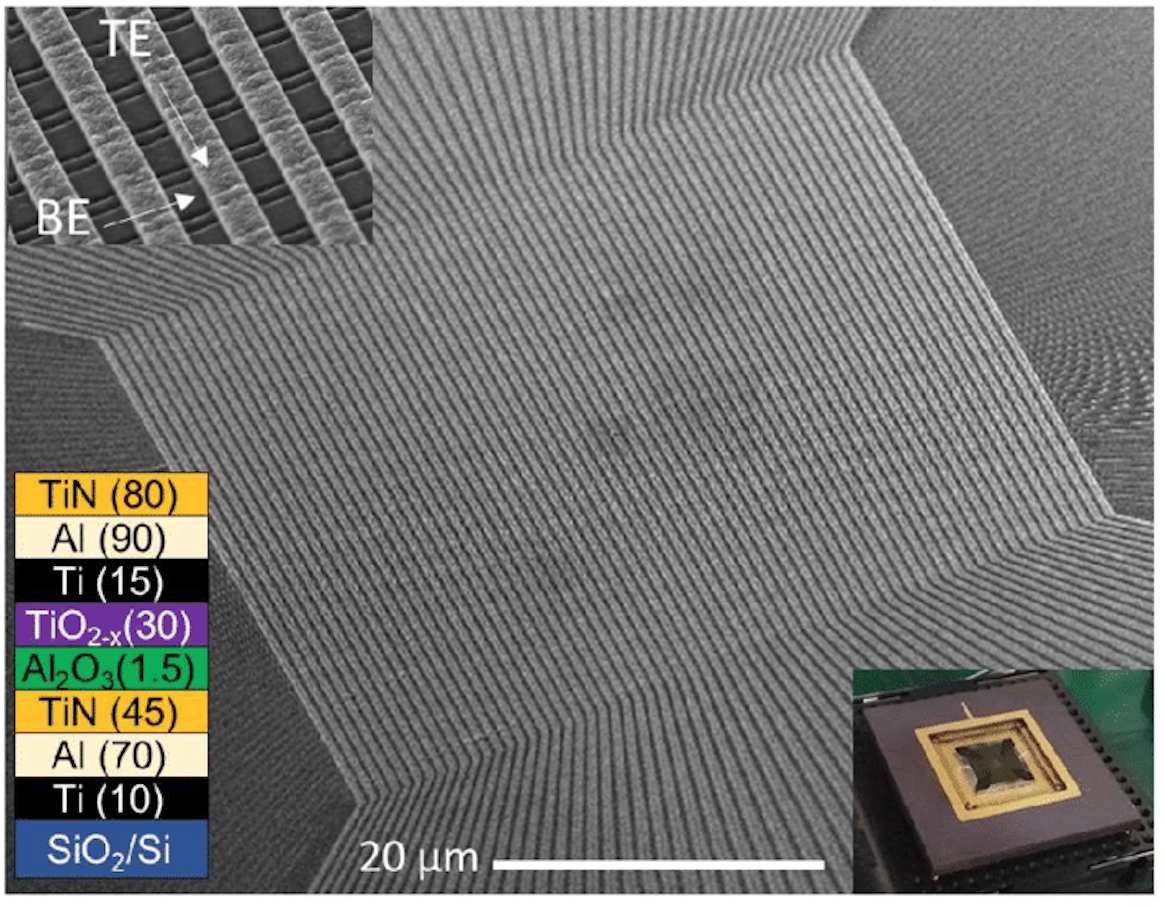}
    \caption{A crossbar array device of TiO$_2$ non-volatile memristors that implement a synaptic neural network. Adapted from Ref.~\onlinecite{kim2021}.}
    \label{Fig:CrossbarMemristors}
\end{figure}

So far it has been easier to implement networks of artificial synapses than neurons. Networks of neurons using new quantum materials have only been reported at the level of numerical model simulations.\cite{nunez2021} A recent example is the work of Oh {\em et al.},\cite{oh2021energy} where a rectified linear unit (ReLU)\cite{agarap2018deep} neural network was simulated from a VO$_2$ neuron device data, predicting an excellent energetic performance. An earlier work in this direction was done by Jerry {\em et al.},\cite{jerry2017ultra} where the stochastic properties of VO$_2$ devices were simulated to perform a digit recognition task.  This focus on synaptic networks is due to several reasons. First, non-volatile resistive switching, relevant to synapses, has been intensively investigated over the last 20 years.  Furthermore, non-volatile effects are easier to observe since they occur in an astonishing number of transition metal oxides.\cite{aonowaser2007,wong2012} Importantly, they already shows excellent performance in devices made with simple oxides, such as TiO$_2$, TaO$_2$, HfO$_2$,  etc.\cite{chen2009} For example, a recent synaptic neural network implemented with TiO$_2$ is shown in Fig.~\ref{Fig:CrossbarMemristors}, which illustrates the current state of the art.

In contrast, the progress in networks of artificial spiking neurons has been slower. Implementing such devices requires quantum materials that exhibit a stimulus-driven insulator to metal transition, such as the vanadates and nickelates mentioned above. These materials are harder to fabricate in good quality thin films, as the transition properties strongly depend on the substrate and deposition conditions.\cite{hank2021}

It is a pressing issue to make progress in the implementation of networks of neurons. A key difference between neurons and synapses, beyond their type of resistive switching, is a functional one. The essential function of a synapse is to encode the coupling intensity between neurons, but synapses do not necessarily need to interact. This is in stark contrast to neurons, where the excitation state of an upstream one needs to elicit, or at least contribute to, the excitation of several downstream. 

\paragraph*{Neuron-neuron interaction.}

Achieving control and physical insight into neuron-neuron interactions is a significant challenge. Almost all the work in the field has remained at the level of a single neuron device\cite{stoliar2017,yi2018biological,delvalle2018}. This includes the work of Jerry {\em et al.}\cite{jerry2017ultra}, where a multi-neuron VO$_2$ device was implemented. However, the neurons were independent, not interacting, and the device function exploited the stochastic behavior of the ensemble of independent neurons. 

A first step to achieve the neuron-neuron interaction would be to get one excited neuron to induce the excitation state of another. It should be kept in mind that the interaction should be a priori scalable to multiple neurons, since building networks is the ultimate goal. Since Mott neurons rely on an electro-thermal incubation process\cite{rocco2022} we envision different types of neuron-neuron interactions based on both electric and thermal coupling.

\paragraph*{Electric coupling.}

The simplest demonstration of electric coupling is to connect two consecutive Mott neurons so the excitation of the first one induces the excitation of the second. Shriram et al.\cite{shriram2018} implemented such a monosynaptic neuron circuit. The thyristor,a conventional semiconductor electronics component,\cite{rozenberg2019ultra} which was recently recognized to have memristive properties that are qualitative similar to Mott materials, demonstrated coupling with leaky-integrate-and-fire artificial neurons.\cite{stoliar2021jeffress,stoliar2021}

Another way to implement electric coupling is capacitively. The simplest spiking neurons based on Mott materials are spiking oscillators, where a Mott device is connected in parallel to a capacitor.\cite{pickett2012neuristor} Initially, the latter charges while the former is in its natural insulating state; then, when the voltage $V_C$ reaches a threshold the Mott device undergoes a metal-insulator transition leading to a spike of current as the capacitor discharges; and then the cycle restarts. This behavior was the basis of the neuristor proposed in 2012.\cite{pickett2012neuristor} In more recent work, Adda {\em et al.}\cite{adda2020direct,adda2022} imaged and studied the dynamics of these oscillations in detail. Systems of oscillators can show complex emergent dynamical behavior, as known from the behavior of coupled pendulums,\cite{winfree1967biological} ranging from synchronization to chaotic motion. This establishes a parallel with neural behavior, from brain-waves measured in elctroencphalograms to chaotic brain discharges during epileptic seizures.

\paragraph*{Thermal coupling.}

A second type of coupling is thermal, which exploits the fact that Mott neurons can be triggered by self-heating under significant power injection. del Valle {\em et al.} demonstrated this approach with a caloritronics-based neuristor\cite{delvalle2020caloritronics} in which the heating from a current pulse through a thin wire of Ti in close physical contact with a VO$_2$ device induced resistive collapse of the insulating state. An important step would be to replace the Ti wire by a Mott device, where the heat produced by driving one Mott neuron would induce the excitation ({\em i.e.}, resistive collapse of a second Mott device. The viability of this type of coupling iwould open interesting possibilities to explore by arranging the Mott devices in different geometries, which may enable spatial spike train propagation via a heat wave propagating through the devices.

\begin{figure}[t]
    \includegraphics[width=\columnwidth]{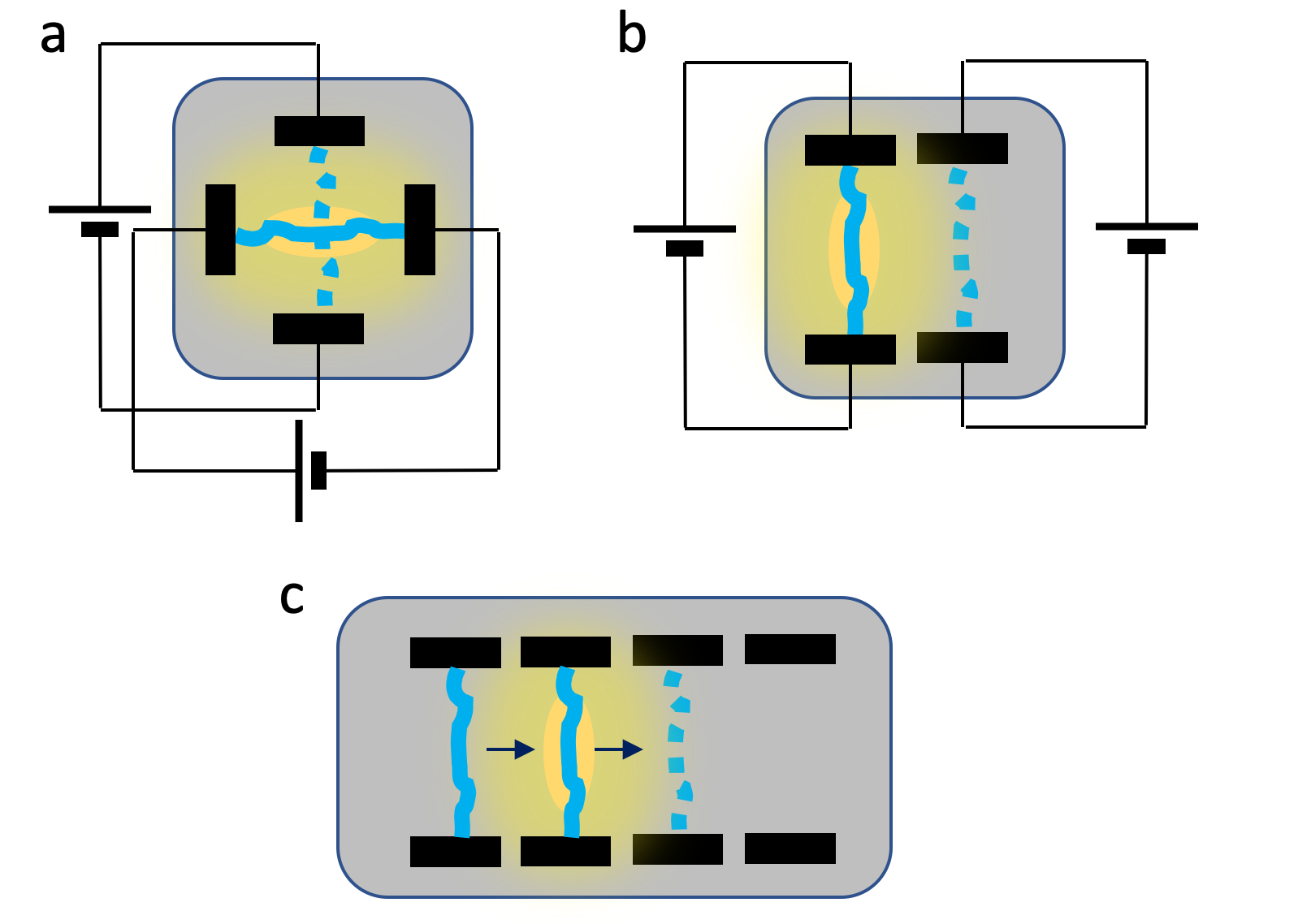}
    \caption{Different geometries for thermal coupling. The parallel geometry may be extended to produce spike trains though a propagating heat wave.}
    \label{Fig:ThermalCoupling}
\end{figure}

\paragraph*{Flux quantization and dynamics.}


A very different approach towards complex networks with cerebral plasticity, is provided by a random  Josephson network made up from the perovskite YBa$_2$Cu$_3$O$_7$.\cite{cybart2009very} Such a random Josephson network has a very high speed response from the stimulus of a train of input spikes and is capable of short-term learning. The output is a series of voltage spikes representing the memory state of the Josephson circuit.  

A closed superconducting loop can maintain a circulating current and trap magnetic flux in units of flux quanta in the loop.\cite{tinkham2004introduction} For appropriate loop parameters ($LI_c \approx F_0$) a single flux quantum either positive or negative, can be trapped.  Furthermore, if the superconducting loop encompasses a Josephson junction, upon reaching the critical current of the junction, a single flux quantum will enter or exit the loop. This technology has been applied to logic circuits, the simplest form being a flux shuttle.\cite{fulton1973flux,likharev1991rsfq} The switching time is $\approx 10^{-12}$~s and the energy consumed in that switch is on the scale of attojoules.

\begin{figure}[t]
    \includegraphics[width=\columnwidth]{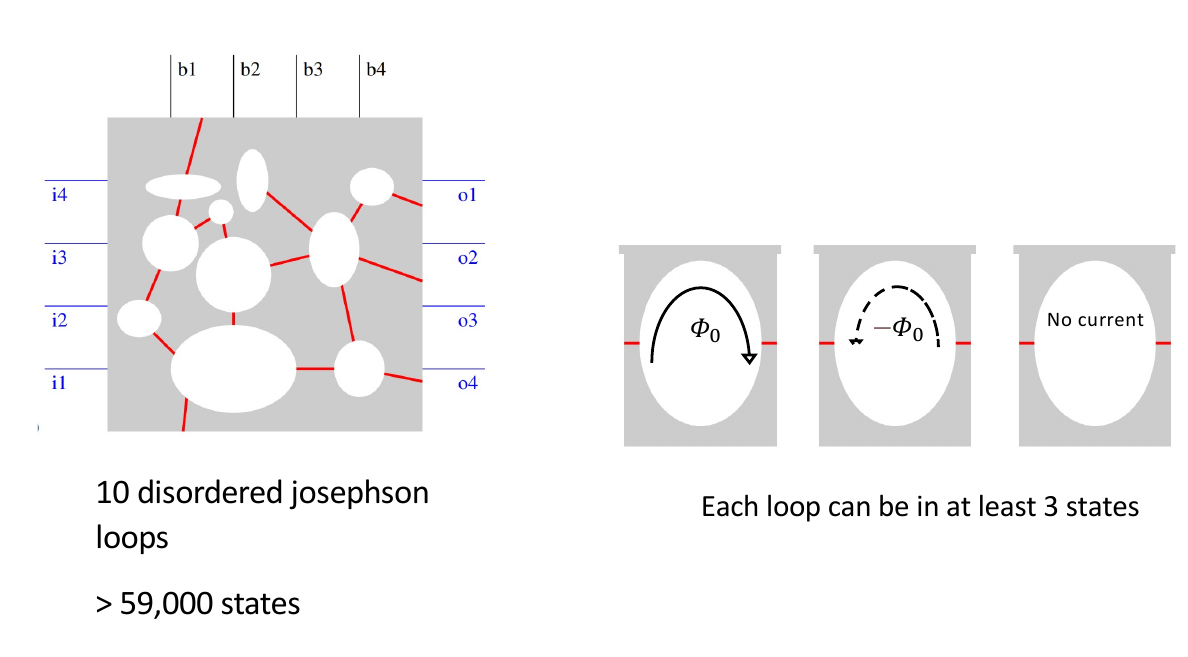}
    \caption{Each loop can be in at least 3 flux states. Flux can be moved in or out via a Josephson junction (red bars). To drive and probe  the 10 loop memory, inputs ($i_j$) outputs ($o_j$) and feedback ($b_j$) are used. 10 disordered loops can have at least $3^{10}$ states.}
    \label{Fig:SCloops}
\end{figure}

Fully coupled randomly disordered superconducting networks with open-ended channels for inputs and outputs are a new architecture for neuromorphic computing.  We have shown that such a network can be designed around a disordered array of synaptic networks using superconducting devices and circuits as an example.\cite{goteti2021superconducting} A similar architectural approach may be compatible with several other materials and devices. The randomness is important as it allows scalability at an exponential rate providing substantial computing power and memory. The scalability of such a architecture is illustrated in Fig.~\ref{Fig:SCloops} where the number of states available in a disordered array is $3^n$ for an array of $n$ loops that allow only a single flux quantum. In the figure showing a network of 10 loops, the number of possible states is $3^{10}$.  

A simple array of three multiply coupled (interconnected) disordered loops containing Josephson junctions forms a fully recurrent network together with compatible neuron-like elements and feedback loops, enabling unsupervised learning.\cite{goteti2021superconducting} Several of these individual neural network arrays can be coupled together in a disordered way to form a hierarchical architecture of recurrent neural networks that is similar to that of a biological brain. The plasticity of this structure can be built into the design by modifying the number of fluxoids trapped in the loop and the binding energy (the $I_c$ of the Josephson junctions. 

Even more flexibility has been suggested\cite{goteti2021low} by a hybrid design of a) the array described above and b) a more “stable” platform like Hydrogen doped RNiO$_3$.\cite{shi2014colossal} It has been shown in nickelates (RNiO$_3$) that H$^+$ ions can be introduced and by controlling electric field, the ion H+ will drift towards a negatively charged electrode resulting in a much higher resistance along that pathway, at time scales that are at the other end of the spectrum and complimentary to those of random Josephson junction arrays. This was accomplished via a series of voltage pulses and resulted in a long-term memory. The substantial differences in time scale between the Josephson array, which is driven into a particular neuromorphic state, and the H-doped nickelate, which is also driven into a neuromorphic state by voltage pulses, allows a large time-spectrum, which opens the opportunity for tuning neuroplasticity in a large range of time scales.

\subsection{\label{Sec:SpinCurrents}Coupling magnetic devices}


Section~\ref{Sec:MagnetizationDynamics} describes different magnetic devices that can function as neurons and synapses. For these to accomplish useful cognitive computing, large numbers of them need to be coupled together with controllable coupling strengths. Just as magnetic devices have a wide variety of behaviors, they have a wide variety of ways to couple them, including electrical, electromagnetic, and direct exchange coupling.

The read-out mechanism of spintronic devices is generally either magnetoresistance of some form or generalized Hall conductance. The field of spintronics blossomed with the discovery of giant and tunneling magnetoresistance,\cite{baibich1988giant,binasch1989enhanced,parkin2004giant,yuasa2004giant} the large change in resistance on changes in the magnetic configuration. Such changes in resistance also provide an electrical mechanism to couple devices because changes in the current through these devices change the current through subsequent devices, thereby changing the spin-transfer or spin-orbit torques on those subsequent devices.\cite{romera2018vowel} The efficiency of this approach is determined by both the magnetoresistance of the devices and the efficiency of the spin-transfer torques that change when the current and voltage change. This electrical approach to coupling has the advantages that its range is not limited and its signals can be readily integrated with CMOS circuitry to aid in amplification and fan out.\cite{prasad2021associative} It has the disadvantage that the ohmic losses due to the current can limit its energy efficiency.

A coupling mechanism that does not suffer from ohmic losses is magnetostatic coupling between the magnetizations of magnetic devices. Such coupling is exemplified in the use of artificial spin ice\cite{wang2006artificial,nisoli2013colloquium} to implement coupled systems of Ising spins. An artificial spin ice consists of an array of nanomagnetic structures that interact through the magnetic fields that they all possess. The nanomagnets are typically stable in one of two configurations, giving the two states of an Ising model. Since the beginning of artificial intelligence development, Ising spin systems have served as a model for artificial neural networks. Spin switching between two states indeed mimics neuron spiking, while the strengths of coupling between spins emulate synaptic weights. The famous Hopfield model\cite{hopfield1982neural} shows for example how memories can be stored in such a spin system, in an associative way that resembles the brain's behavior. 
This process can exploit thermally activated transitions as in Boltzmann machines.\cite{hinton1986learning} Building hardware Ising spintronic systems to implement neural networks natively is thus a compelling solution to compute through the physics of coupled nanomagnets.

Spin ice is a natural platform for such purpose. Recent proposals show that assemblies of coupled nanomagnets can compute through energy minimization or through their transient behavior via reservoir computing. Their reconfigurability by removing nanomagnets elements has been demonstrated in stochastic Kagome lattice imitating nano-Galton boards.\cite{sanz2021tunable} The coupling between magnets in these systems is local by nature. It is therefore a challenge to build more powerful, state-of-the-art, all to all connected neural networks in which each magnet is coupled to each other magnet in the network in a reconfigurable way. For artificial intelligence applications, the efficiency of hardware Ising machines such as quantum annealing systems that couple Josephson junctions, or CMOS and optical implementations is reduced by the limited number of neurons that can be all connected to each other. 

\begin{figure}[t]
    \includegraphics[width=\columnwidth]{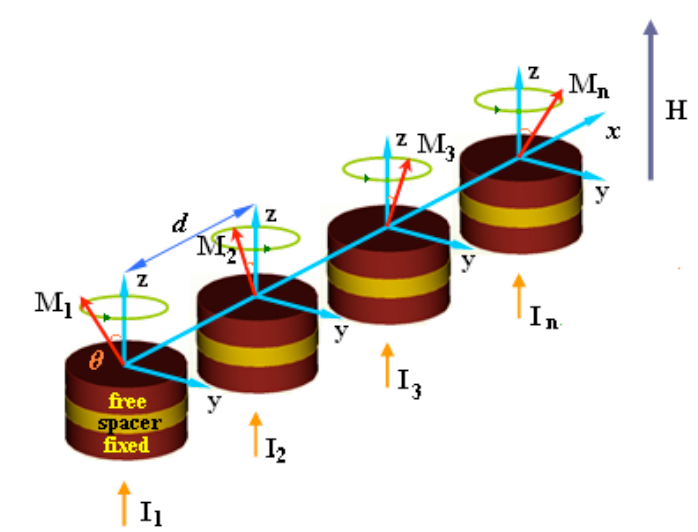}
    \caption{\textbf{Coupled spin torque oscillator chain.} Long chain of spin torque oscillators with perpendicular anisotropy with the following soft layer parameters: $D=60$ nm, $t=1$ nm, $M_s=1000$ emu/cm$^3$, $K_{u,h}=1$ kerg/cm$^3$, $\alpha=0.01$.  An external perpendicular magnetic field of $4$ kOe is applied. The current has a random distribution of 2~\% between different spin torque oscillator (similar effects are obtained for distributions in other parameters).}
    \label{Fig:VL2}
\end{figure}

Magnetostatic interactions also couple spin-torque oscillators, although these oscillations are at gigahertz frequencies and so can also be thought of as RF coupling. Similarly if the ferromagnetic material is continuous between two oscillators, spin waves can couple them. There are some similarities and difference between these two types of coupling.\cite{Awad2015,Kendziorczyk2014,Houshang2016,Slavin2006} The strength of magnetostatic interactions decreases rapidly with distance and they do not introduce any phase shifts in the interactions between spin-torque oscillators. As a result, for a spin-torque-oscillator array, magnetostatic interactions lead to near-neighbor coupling that propagates as a chain. As an example, a frequency-synchronized linear array with random parameter distributions but with a consecutive angle shift between adjacent spin-torque oscillators results in an angle wave along the array (Fig.~\ref{Fig:VL2}).\cite{volvach2021micromagnetic} On the other hand, spin wave interactions between spin-torque oscillators introduce phase differences due to the finite spin wave velocity and their strength may decay slower than that of magnetostatic interactions. \cite{slavin2009nonlinear}

Fortunately, spintronics offers multiple ways to connect magnetic devices through long range interactions. Optical and spin wave beams could be exploited for this purpose. Another promising option is the wireless coupling offered by microwave signals. Spin-torque nano-oscillators can take a dc electrical input, and emit a microwave power that is a non-linear function of the input, resembling the ReLu neural activation used today in artificial neural network.\cite{nair2010rectified} This radio-frequency signal can then be broadcast to other elements of the network, synapses and neurons. Spintronic diodes can natively sense this microwave signal and convert it to a dc voltage that can be used to feed other devices.\cite{tulapurkar2005spin,sankey2006spin-transfer} Recent experiments show that these spintronic diodes perform the elementary synaptic operations and that their strength can be controlled in a non-volatile way when they are interfaced with oxide materials that undergo a phase change near room temperature such as vanadium oxides.\cite{Xu2021} 

The potential of magnetic tunnel junctions for neuromorphic computing has been demonstrated by leveraging the non-linear dynamics of harmonic, sinusoidal oscillations induced through a dc spin-polarized current.\cite{Torrejon2017neuromorphic} However, a key ingredient of biological neurons is to emit spikes when the membrane potential overcomes a threshold. Therefore, a major class of bio-inspired neural network algorithms—called spiking neural networks—exploits spikes for computing. Such networks encode information in the timing between spikes in addition to the rate of the spikes. They circumvent the heavy external circuitry that is usually required for training synaptic weights. The electrical spikes applied to the synapses can indeed directly induce weight modification and learning through, for example, the bio-inspired learning rule called spike-timing-dependent plasticity.\cite{bi2001synaptic} The multifunctionality of spintronic devices can be leveraged to achieved spiking behavior. Two methods have been proposed.

The first concept is based on the windmill magnetization dynamics in magnetic tunnel junctions with two weakly coupled free layers.\cite{matsumoto2019chaos} A spin-polarized dc current induces a perpetual switching of both magnetic layers \cite{slonczewski1996current}. Neuron-like spikes are obtained by exploiting the transient behavior of this windmill motion in response to voltage pulses. The second way proposed to achieve the spiking behavior is to harness the dynamics in antiferromagnets \cite{khymyn2018ultra,zhang2020antiferromagnet,liu2020synthetic}. The coupled dynamics of the two magnetic sub-systems can indeed be rewritten as an equation similar to the phase dynamics of Josephson junctions, known to exhibit spikes of activity. Recently, a ferromagnetic structure that gives similar behavior in ferromagnetic systems was proposed.\cite{markovic_easy-plane_2021} These proposals are theoretical for now and demonstrating them experimentally is an important challenge for future spintronic neuromorphic systems.

One of the most useful features of strongly correlated oxides is that the ground state properties can be modified easily by various means, including implantation of or irradiation with light, inert ions like He and Ar.\cite{Chen1989,Lang2010,White1988,Naitou2016,Stanford2016} A key advantage of this approach is the ability to focus a beam of light ions down to a beam diameter on the order of a few nanometers, localizing the effect the light ions have on the correlated material.\cite{Naitou2016,Stanford2016} One application is modifying the oxygen stoichiometry within a narrow region of a few nanometers. Since the properties of oxides are closely linked to their oxidation states, such modifications give control of material properties with exquisite lateral resolution. 

\begin{figure}[t]
    \includegraphics[width=\columnwidth]{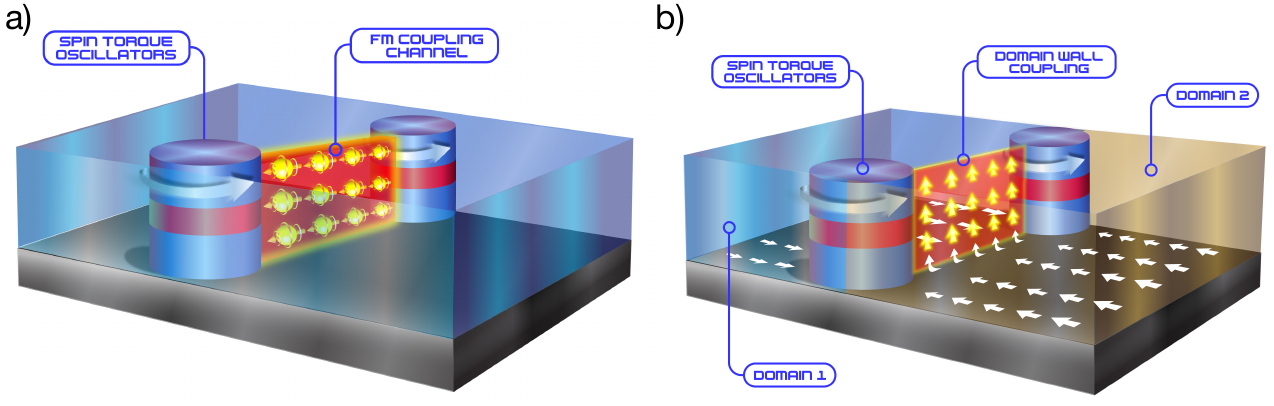}
    \caption{(a) A He beam could be used to write a ferromagnetic (FM) channel within a paramagnetic film that acts as a coupling channel between two spin torque oscillators. (b) Similarly, a domain wall of nm width can be drawn between to oscillators to allow for coupling.}
    \label{Fig:AF1}
\end{figure}

Rare earth nickelates, which have a metal-insulator transition coupled with a paramagnetic-
antiferromagnetic transition\cite{Das2021} provide properties that can be dramatically modified. Upon irradiating the films homogeneously with a beam of helium ions, both transitions are strongly suppressed, and the resulting ground state is a metallic/paramagnetic material. Measurements indicate a sizeable change in the nickel valency state as the driving mechanism behind the suppression of the transition. Similar effects are expected to occur in other oxides such as manganates, titanates, vanadates, etc., which feature numerous magnetic ground states across phase diagrams where the oxidation states are the control parameter. Furthermore, a focused ion beam could be used to draw domains of paramagnetic regions within a backdrop of antiferromagnetic regions (or vice versa). This could be extended to delineate magnetic domain walls along any desired pattern. Finally, these domain walls could be drawn to connect two magnetic oscillators, thus potentially resulting in coupling them through magnons or spinons that live exclusively on the domain wall (Fig.~\ref{Fig:AF1}). The tunable properties of correlated oxides allow versatile ways of designing domain walls with different materials, and could then be used to create and control new coupling mechanisms.

Spintronic structures for neuromorphic applications often involve spin textures that can be characterized as topological objects, such as vortex cores and skyrmions.\cite{Zang2018} Such topological objects possess a winding number, given by: $W=\frac{1}{4\pi}\int{\hat{m} \cdot (\partial_x \hat{m} \times \partial_y \hat{m})dxdy}$. $W$ is the winding number and $m$ is the unit vector of magnetization and two topological objects are said to have distinct topologies if their winding numbers are different. A topological object can exist only under specific conditions, and therefore modulating its local envirenment can provide a new form of tunability. 

It was recently realized~\cite{Arava2021} that the stray field environment from four Permalloy nanomagnets surrounding a Permalloy disk can be described using topological arguments and a discretized form of the winding number, given by $W=\frac{1}{2\pi}\sum \beta \Delta \theta_i$, where $\beta$ is the relative change in orientation of the nearest neighbor’s magnetization ($\pm 1$), and $\Delta \theta_i$ is the angular difference between the magnetization of two nearest neighbors. It was found that the winding number of the topological spin texture (vortex, antivortex, or uniform) in the permalloy disk would always match the winding number from the stray field in the surrounding nanomagnets. A similar topological argument was also found to hold true for disks physically connected by exchange-mediated nanomagnets. 

\begin{figure}[t]
    \includegraphics[width=\columnwidth]{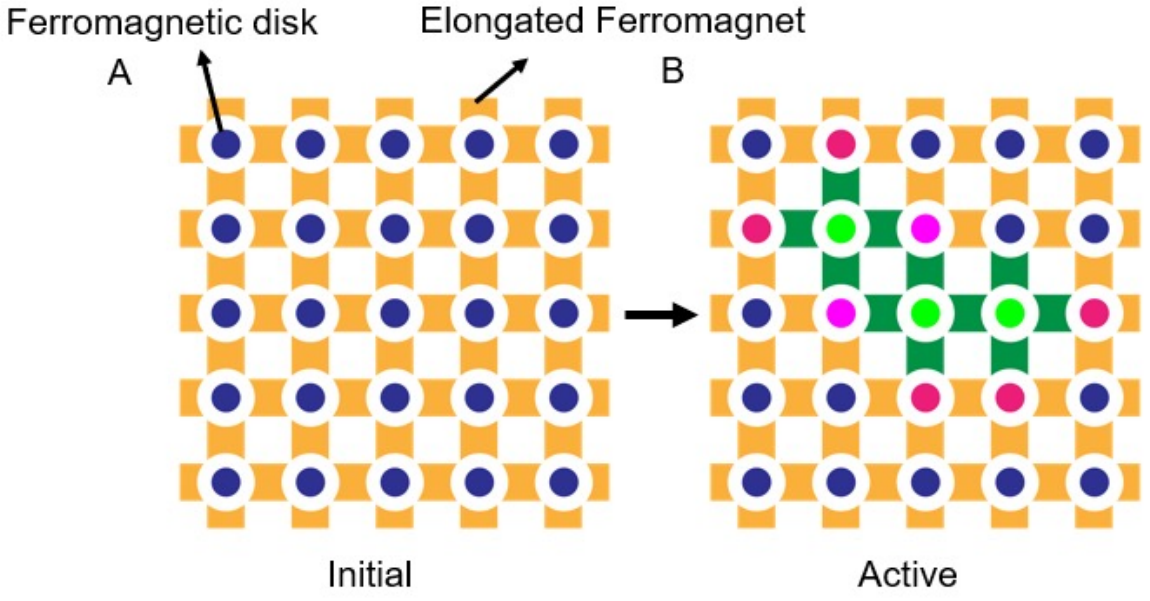}
    \caption{(a) Array of ferromagnetic disks (blue) surrounded by elongated nanomagnets (orange), in which the nanomagnet stray fields provide a topological environment that connects the disks. (b) An active state, for example set by an external stimulus, in which the array can develop regions with distinct topology (indicated by different colors).}
    \label{Fig:APL}
\end{figure}

An array of disk/nanomagnet structures (see Fig.~\ref{Fig:APL}) controlled by an effective topology could thus potentially be implemented in neuromorphic or unconventional computing schemes. An active state (see Fig.~\ref{Fig:APL}) described by regions with different topologies, indicated by the blue or the green disks, could then be set, for example by an external field or current, and probed, ({\em e.g.},  by resistance measurement) to distinguish different topological states of the array. We foresee the ideas described here being extended to spin textures such as skyrmions, in which the arrays would no longer be planar (such as in Fig.~\ref{Fig:APL}), but could have a three-dimensional structure. 

\section{\label{Sec:Outlook}Outlook and Perspective}

In this paper, we have presented some of the recent remarkable progress in developing quantum materials that enable novel devices, developing these devices, and identifying networks that can take advantage of the properties of these devices. Despite this progress, there is much more to be done before these materials become a commercial reality. In this section, we describe a few of the outstanding directions for future research for devices, networks, and the related technologies of sensing and harvesting.

\subsection{Perspectives for individual neuromorphic quantum devices}

In spite of the wide range of properties possessed by quantum materials, the constraints provided by designing competitive devices for neuromorphic computing are equally tight.  The search for the optimal materials is quite difficult. There are several routes for further progress.

One route is to embrace computational techniques to speed up the search for materials with desired properties. Computational techniques have been an invaluable tool in the study of quantum materials for neuromorphic computing. Beyond merely an investigative tool, such techniques, particularly in combination with state-of-the-art machine learning algorithms, have the potential to transform quantum materials discovery and design as well. While electronic structure calculations based on density functional theory provide “clean” data (free of the vagaries of experimental measurement) on the intrinsic properties of materials, machine learning algorithms can unveil the underlying structure-property relationships and predict where the next breakthrough material will be found. The challenge is developing a sufficiently large dataset to build accurate machine learning models for quantum material property predictions 

Another way to design optimal materials is to go beyond the search for one specific material that has all the desired properties, but to combine a set of materials with all said properties instead. The physical properties of strongly correlated materials appeal to a broad scientific community because of the versatility and tunability of their electronic responses via internal and/or external perturbations.\cite{schlom2007strain,imada1998metal,ngai2014correlated} However, the number of ways to control a single correlated material is limited by the available internal degrees of freedom. This presents a difficult challenge when a device concept requires a specific mechanism of control that is not accessible within a material. A general solution to this limitation is to judiciously design heterostructures that hybridizes the functionalities of two seemingly unrelated materials. In this fashion the properties of one material can be used to change the functionalities of the other. 
This could provide exquisite control of the magnetic materials by voltage application, modification of the metal-to-insulator transition via optical means, changes in the optical properties of semiconductors by electric fields, or control of the conductivity with stress.

Artificial spiking neurons based on Mott insulator materials are a clear example of how much work is still needed to understand the properties of these materials in which novel devices are based. Significant progress has been recently made towards understanding of key physical phenomenon that enables it, namely, the unique non-destructive and "self-healing" electric breakdown in a strongly correlated insulator. Still, questions remain on how to better understand the nanoscale thermodynamics of the device. This is relevant because at those small dimensions both, heating and cooling occurs fast, even within nanoseconds.
Moreover, better understanding of these thermodynamics is also needed as the Mott materials are driven significantly out of equilibrium, producing the growth of filaments in the conductive phase, that are believed to undergo an insulator to metal transition either driven by self-heating or carrier injection,\cite{kalcheim2020non} eventually having to relax from the metastable phase.\cite{tesler2018relaxation,delvalle2020caloritronics}

Carrier recombination dynamics as the metallic state decays back to the insulating state is another factor that can pose fundamental limits to the operating frequencies of Mott threshold switching neurons.\cite{shi2021dynamics} These challenges for the Mott spiking neuron bring to light another important issue in the development of novel materials and devices –- the ability to measure their behavior at the nanoscale. In order to characterize the material properties in different phases, it is necessary to use different analysis techniques such as x-ray diffraction, Raman spectroscopy, Terahertz transmission, etc. Ideally, these measurements would be made in operando, particularly in materials undergoing phase transitions that give them their desired properties. A pump-probe characterization method is ideal for this purpose, since it can use the different source combinations for  the driving mechanism and analysis method. For example, vanadium oxides can be triggered with external voltage, current source, or thermal source and the formation of filament can be optically mapped such as in-situ X-ray nanoimaging.\cite{shabalin2020nanoscale,adda2020direct} 

It is not a trivial task to integrate the driving excitation system for the phase transition and the analysis tools. When using a Raman spectrometer to analyze a temperature-driven device, careful design of the thermal chamber for the sample is necessary. In addition, while driving the phase transition with one source, the probe might also contribute to the phase transition so that it is important to measure the dependence of the signal on the probe power.\cite{miao2018phase,abreu2017ultrafast} The design can be more sophisticated when triggering the phase transition with Simultaneous stimuli (thermal, electrical, and optical) or characterizing the materials with some sophisticated techniques. 

Achieving high spatial resolution for mapping filament formation in phase transition materials, tip enhanced Raman spectroscopy seems to be a promising approach. However, both the optical probe and strong electrical field could induce the phase transition in addition to the original excitation source. The designed optical beam spatial profile, the applied electrical field, and the size of the tips determine the accuracy of the results. While considering optical characterization of the dynamics of phase transition with high temporal resolution and the behavior over a specific band, the use of broadband light source or ultrafast laser is inevitable. Consequently, careful design of optical system for aberration compensation is also necessary. Integrating the characterization system with the driving system for phase transition requires not only a careful design of the individual system but also the investigation of the contribution of multiple driving sources to the phase transition.

A key characteristic of neuromorphic devices is their high speed, with operation times that can be well below the nanosecond time scale. Brain components, on the other hand, operate much slower, at the millisecond time scale. The speed of artificial neuronic devices is an advantage as it can lower the overall energy consumption of a system. Furthermore, since massive parallelism of the brain is challenging to achieve in hardware, having high speed individual components may compensate for a lower degree of intrinsic parallelism by time multiplexing, to reach an overall equivalent operating time at the system level. 

While typical computing always evolves toward doing the calculation faster, it may also be important for neuromorphic computing to optimize materials toward slower performance or performance over mixed time scales. The brain’s functionality may indeed originate from its processing speed matching the input rate of the information it is processing. Consider, for instance, the task of auditory or visual pattern recognition which the brain can accomplish with very high accuracy albeit slow speeds. Recent work has shown various ways to accomplish this using neuromorphic hardware and software,\cite{romera2018vowel,oh2021energy} which can be processed at speeds far faster than the brain. These tantalizing results show how some processes may result to be faster than in the brain. 


\subsection{Building networks: connectivity, dynamics and emergence}

As shown in this review, there has been considerable progress in designing quantum materials that emulate individual elements for neuromorphic computing: oscillating or spiking neurons, and diverse types of synapses with short- or long-term memory. There has also been progress in coupling a small number of devices together to make networks capable of carrying out calculations. A big challenge is to demonstrate that the networks based on the novel devices are competitive in terms of speed, size, resilience, energy efficiency, and production costs. Demonstrating capabilities requires substantial effort particularly for devices and networks that are not based on CMOS substrates because of the current market domination enjoyed by CMOS.

Demonstrating the advantages of a device based on novel materials requires the implementation of a network unless the device is a drop-in replacement for an existing device. In other cases, it is necessary to consider the supporting circuitry. Here, the most straightforward approach is to base the supporting circuitry on CMOS because it has sufficiently established complexity to address almost all tasks. As CMOS runs into the limits of scaling, other possibilities may evolve that avoid some of the constraints created when adopting it. CMOS circuits have been primarily designed for very efficient digital logic, so when the supporting circuitry can be used in that fashion, it may not add much overhead. Digital logic allows for high precision operation with high noise tolerance. Many novel devices aim to use relatively low precision analog encoding of information to take advantage of the fact that most neuromorphic computing tasks do not require high precision. While this approach can lead to low energy use in the device itself, straightforward analog operation using voltages other than the maximum or minimum of the embedding circuitry can lead to significant overhead. Significant design effort is required to ensure that the total energy cost of a new device is competitively small.

One approach to reducing the energy cost is to pulse the voltages rather than apply them in steady state. As an example, consider a superparamagnetic tunnel junction used to generate random bits. It can be much more efficient to read its state with pulses using a pre-charge sense amplifier than to read it continuously.\cite{mizrahi2018neural, ababei_neuromorphic_2021} In addition, the details of the supporting circuitry require us to consider what application the novel device will be used for. Designing this circuitry to carry out tasks for an existing approach is the best way to achieve a direct comparison of the novel device’s efficiency.\cite{daniels2020energy,oh2021energy}  

A drawback of using CMOS as a substrate is that it uses a relatively fixed window of operating voltages.\cite{weste2010cmos} Devices that need larger voltages will require specialized circuitry to apply the larger voltages, even if only for the forming step and will require high power transistors to control that voltage. Such circuits will greatly increase the costs in circuit area and energy consumption for those devices. On the other hand, devices that operate at lower voltages than that of the CMOS circuit do not provide equivalent savings. Either the voltage would need to be brought down through control transistors, the resistance and energy dissipation of which might negate any energy savings from lower operating voltages, or additional low voltage circuitry would need to be introduced, greatly increasing the needed circuit area. These issues highlight one desirable property to guide the search for device materials intended to operate on a CMOS substrate -- that they operate efficiently at voltages close to those of the CMOS circuitry in which they will be embedded. Alternatively, designing novel CMOS circuits optimized for the new devices could be justified if the advantages offered by a novel device is sufficiently great. 

Another challenge for building artificial neural networks and highly connected systems from quantum materials are the large spatial variations in heating and temperature. These may become important when massive amounts of time-dependent signals are used to process and encode data. Next-generation computing technologies operate at exaFLOPS ($10^{18}$ floating-point operations per second).\cite{ashraf2020empirical} CMOS-based von Neumann-based architectures of such a complexity would consume approximately 20~MW to 30~MW of power, roughly the electrical energy of a whole city. Bio-inspired computation could have much lower power consumption using integrated non-volatile memory and logic and exploiting their learning capabilities from unstructured data. 

As an example, local thermal dissipation in the commonly used crossbar architecture may cause unwanted interference between artificial synapses which are in close physical proximity. Moreover, temporal coincidences of signals may also cause local thermal disruptions and perhaps cannot be ignored when systems are packaged at nanoscale-dimensions and in three dimensions. In traditional two-dimensional architectures, thermal management is facilitated by the fact that the neural network is proximal to a large thermal sink ({\em i.e.}, the substrate). However, a three-dimensional system can no longer rely on this thermal sink. Artificial neural networks based on quantum materials therefore raise the challenge of exploiting temperature changes for computing while maintaining thermal variations in the range in which the system can operate.

Low temperature operations provide a higher efficiency use of a wider set of quantum materials. These operations come with different energy efficiency issues than room temperature.  Superconducting circuitry operates at extremely low voltages so that novel devices in superconducting circuits will also need to operate at low voltages.\cite{holmes2013energy} Low temperature CMOS is also in development and may provide some of the control circuitry for superconducting circuits.  While it will be more efficient than room temperature circuits, it will be much less efficient than superconducting circuits and its energy consumption. This energy consumption will have to be accounted for to the extent needed for such circuit coupling in room temperature environments. It also requires impedance matching with superconducting circuits,\cite{ mccaughan2019superconducting} which carries additional overhead. Finally, for low temperature applications, any heat generated will have to be transferred to room temperature, which uses 75 times more energy in the ideal case (Carnot efficiency) and closer to 1000 in practice.\cite{holmes2013energy} Still, even with the overhead, superconducting circuits can be more efficient in applications like those more compute intensive than data intensive.

One device that we have featured in this review illustrates some of the issues associated with demonstrating viability. Despite much progress at the device level, there is only limited development of the Mott neuron\cite{stoliar2017, yi2018biological, delvalle2020caloritronics} and the neuristor\cite{pickett2012neuristor}, into networks. Establishing the utility of these devices requires progress on the significant challenge of interconnecting these neurons and understanding how they couple and interact when driven out-of-equilibrium. Many paths hold possibilities. For instance, thermal coupling was recently demonstrated in the spike firing by heating of a metallic Ti nanowire crossing the gap of a VO$_2$ device.\cite{delvalle2018}. Another possibility derives from the interaction in the filamentary incubation times of two Mott neurons in physical proximity. This coupling raises the possibility of different multi-electrode geometries with possible thermal and electro-thermal coupling. The viability of these devices starts with being able to implement a system of two, or even many Mott neurons, where the firing of an upstream neuron can drive the spiking of a downstream one. Making a useful network will require implementing synaptic tuning, that is making non-volatile changes in a resistive switching material driven by the spiking. 


One approach to neuromorphic computing that might eliminate much of the CMOS circuitry is to use optics. A key component of such implementations might be phase transition materials. While  phase transition materials have been attracting research attention due to their hysteresis behavior, {\em i.e.}, memory functionality, and significant variation of material properties in different phases,\cite{abreu2017ultrafast} they are also optically active. The phase transition can be driven by various stimuli. 
Analogously to the conductivity change in electrically driven phase transition materials, the variation of optical properties\cite{wu2014microwave} allows the development of a photonic-based neuromorphic system that takes advantage of optical connectivity for scalability and speed to increase the throughput of such systems. 

Vanadium dioxide can have more than a 15~\% change in reflectivity, more than a 100~\% change in third order nonlinear coefficient, and a less than 100~fs transition time when phase transition is triggered by ultrafast laser pulses. The details of excitation beams such as pulse width, pulse duration, repetition rate, wavelength ({\em i.e.}, optical carrier frequency), and polarizations prescribe the optical properties needed in materials.\cite{zhang2019designing} 

Based on variations of optical properties, a hybrid photonic-based computing circuit which consisted of a 2D optical waveguide integrated with phase transition materials deposited on top of it has been implemented.\cite{feldmann2017calculating} To implement the leaky- integrate-and-fire behavior of a neuron, an all-optical circuit with a semiconductor optical amplifier and nonlinear fibers was demonstrated.\cite{kravtsov2011ultrafast} Another potential approach can be explored by using a gain medium as an integrator and phase transition materials as output couplers, since the reflectivity of phase transition material changes with varying incident optical pump powers. To emulate synaptic plasticity, devices with electrically-driven phase transition materials have demonstrated the capability of updating the synaptic weight and generating different output signals in response to different input signals. 


Spintronics also offers promising solutions to enhance the connectivity of neuromorphic networks. The properties of magnetic materials are sensitive to strain, light, electric and magnetic waves, in the dc and rf ranges. By themselves, or combined with other materials, they can generate waves that can then propagate to other magnetic elements, such as {\em e.g.}, spin waves and radio frequency signals. Wave-like connectivity between neuromorphic spintronic elements can be exploited to build densely connected neural networks.\cite{Leroux2021radio} The emission and propagation of magnetic quasiparticles in two- or three-dimensional magnetic networks is also an interesting option to achieve high connectivity. 

An intriguing option to enhance the size and connectivity of artificial neural networks is to use interconnected qubits for machine learning.\cite{markovic2020quantum} The advantage of quantum neural networks is the possibility of encoding neurons in the basis states of a quantum ensemble rather than in qubits, thus obtaining neural networks of significant (exponential) size, on a very small number of qubits. A small number of noisy qubits can be made useful for computing by adopting a hybrid quantum/classical approach; by offloading some subroutines such as data pre- and post-processing to a classical computer. Since universal quantum computing is still out of reach for currently existing Noisy Intermediate Scale Quantum (NISQ) devices, quantum machine learning and quantum neural networks could provide more immediate applications for these devices.\cite{markovic2020quantum} 

While the road to viable implementation of networks using novel devices has many roadblocks, the potential benefits keep it promising. The human brain is a clear model that more efficient devices and networks exist for many problems than our current implementations. Bioinspired computation encompasses a broad spectrum of new technologies based on energy-efficient biological systems, particularly on the human brain performance. The human brain has 100 billion neurons and 1000 trillion synapses;\cite{niven2016neuronal} assuming that each synapse stores 5 bits per synapse of information,\cite{bartol2015nanoconnectomic} the human brain storage capacity is about 1 petabyte. Biological intelligence results from complex sensory information processing and cognition, so there are evolutive morphological variations of the brain among different species.\cite{hofman2014evolution} This is clearly seen in the mammalian variation of the cerebral cortex,\cite{herculano2012neuronal} the dependence of the brains' operational temperate range with the encephalization quotient for different animals, or the complexity of the cerebral cortex's neural circuitry.\cite{hofman2014evolution}

How does the neural activity in the brain lead to energy-efficient computation? Remarkable advances in network science open up new perspectives to understand this highly complex process unveiling emerging features of low consumption computing. Despite the neural networks transport electrical and chemical signals through an intricate neuronal web, their behavior is governed by universal laws that capture significant emerging features\cite{miller2009power} that could shed light about elementary information processing of active elements (neurons, synapses, dendrites) and structural properties of the brain (neural network, vascular network).

\subsection{Delivering inputs and energy to the network: sensing and harvesting }

One of the central themes in modern solid-state physics is combining materials in order to find new response functionalities that emerge only when two distinct materials come into proximity. Recently, we showed that in a judiciously designed bilayer that contains two materials with no apparent similarity: a photosensitive semiconductor CdS and a strongly correlated Mott insulator, displays such kind of novel functionality.\cite{navarro2021hybrid} While the Mott insulator itself has very little response to electromagnetic radiation (light), when the two are adjacent one can see a dramatic effect of shining light: the Mott metal-insulator transition is completely suppressed. As it pertains to neuromorphic systems, the Mott insulator in question here has also shown to have spiking behavior in the right conditions. Therefore, an adequately designed device that includes the CdS could result in a photosensitive neuron-like system, which could detect light as it simultaneously uses the spiking behavior to encode information about the light. In this fashion, a system similar to the eye/brain sensor could be developed.

Spin-based systems also offer a wide range of possibilities for native sensing of inputs and energy harvesting. Spintronic sensors have been demonstrated to be competitive for magnetic field sensing, molecular sorting, radio-frequency signal analysis, and spin wave detection.\cite{freitas2016spintronic} Spintronic neural network therefore possesses the ability to sense such inputs directly in the analog domain, without energy-costly initial conversion to the digital domain. Spintronic nanodevices can also harvest energy from signals collected in the outside world, such as radio-frequency or heat.\cite{hirohata2020review} Spintronics therefore offers a promising path to build energy efficient physical neural networks that directly sense the input signals they process, and harvest part, or the entirety of the power they need to operate.

Bioelectricity is the main stimulus in beings to regulate the functions of neuron cells, tissues, and organs. Thus, for a long time, electric biasing has been considered as a key component in realizing neuromorphic computing. However, with the advancement of energy harvesting technologies, other sources of energy may also be used in a well-controlled way. If a neuromorphic system can take advantages of clean, renewable, easy access energies, for instance, solar, hydrogen, and geothermal, the systems can be more energy efficient and environmental-friendly. 

VO$_2$ is an archetypal Mott system and can be used to emulate the functions of neurons and synapses.\cite{del2019subthreshold,cheng2021operando} Our preliminary results have shown that, by combining it with photoactive materials, the metal-to-insulator transition temperature of VO$_2$ can change dramatically under sunlight exposure, illustrating a strong photo response.\cite{navarro2021hybrid} This response is largely due to the Mott nature of VO$_2$, where the electron/hole doping originating from the photoactive materials strongly influences its physical properties. 
Indeed, studies have shown that vanadium oxide devices can be used to store energy locally as  concentration cells.\cite{shi2021dynamics} Many questions related to these new applications remain open: (1) how the sunlight alters the functionality of VO$_2$ as neuristors and artificial synapses; (2) the role of thermal heat from the sunlight in the metal-to-insulator transition; and (3) the mechanism that controls the proton doping/reaction in the VO$_{2\pm x}$ family.

\subsection{Learning in materio}

A major challenge for neuromorphic computing is to achieve online learning with high accuracy and low energy consumption. The state-of-the-art algorithm for training neural networks, backpropagation of error, is not hardware friendly. It requires hardware implementations of complex circuits for computing gradients, storing gradients –- necessitating huge memory resources as there is one gradient per synapse  –- and programming the synaptic elements with these gradients. In addition, physical neural networks require a huge number of complex interconnected artificial neurons and synapses. There are currently substantial efforts to design and build such circuits.\cite{zhang2020neuro} An interesting approach is to exploit network dynamics and physics to extract gradients, which avoids directly computing them through chain derivatives.\cite{ernoult2019updates} However, these approaches still require circuits to store and apply gradients. 

A dream for neuromorphic computing is to create physical neural networks that learn intrinsically, through the interplay of synapses and neurons, as the brain does, without the need of cumbersome external circuits. Biologically-inspired learning rules such as spike-timing dependent plasticity can implement such learning, and can be implemented with memristive devices.\cite{wang2020resistive} In that case, neurons emit voltage spikes that directly program the synaptic devices to which they are connected, depending on the timing of spikes. The problem of this local learning rule is that it does not minimize an error at the output of the network, giving poor accuracy on complex tasks. A solution is to add a supervision term to the learning rule, but the procedure to compute this term is quite mathematical and resulting circuits can be complex.\cite{payvand2020error} 

An interesting alternative to train physical neural networks is to look for learning algorithms that are deeply rooted in physics. Hopfield networks,\cite{hopfield1982neural} Boltzmann machines,\cite{ackley1985learning} and Equilibrium Propagation\cite{scellier2017equilibrium} exploit the ability of a physical system that possesses an energy function to relax to equilibrium for learning. The later algorithm is particularly promising because it has been shown to compute gradients equivalent to backpropagation through time, scales to complex image recognition tasks and implements intrinsic learning. It belongs to a class of algorithms developed to understand how the backpropagation of errors could be implemented in the brain, ones that exploit neural dynamics to compute and propagate gradients in the network.\cite{richards2019deep} These algorithms, being designed to train physical networks of synapses and neurons in brain hardware, are particularly promising for creating physical networks that learn intrinsically, in materio, through the physics of quantum neuromorphic devices. 


\section*{Acknowledgments}
The preparation of this manuscript was done through the collective efforts of the members of the Energy Frontier Research Center (EFRC) Quantum Materials for Energy Efficient Neuromorphic Computing, funded by the U.S. Department of Energy (DOE), Office of Science, Basic Energy Sciences (BES) under Award \# DE-SC0019273.  Figures~\ref{Fig:correlated} and \ref{Fig:AF1} were designed by Mario Rojas Grave De Peralta

%

\end{document}